\pdfoutput=1
\documentclass[aps,prd,twocolumn,superscriptaddress,preprintnumbers,floatfix,nofootinbib]{revtex4-1}

\usepackage{graphicx}
\usepackage{amsmath}
\usepackage[caption=false]{subfig}
\usepackage{siunitx}
\usepackage{placeins}
\usepackage{color}
\usepackage{standalone}
\usepackage{dcolumn}
\usepackage{tensor}
\usepackage{bm}
\usepackage{microtype}
\usepackage{etoolbox}
\usepackage{amssymb}
\usepackage{mathrsfs}
\usepackage{accents}
\usepackage[normalem]{ulem}
\usepackage[dvipsnames]{xcolor}
\usepackage[colorlinks,urlcolor=NavyBlue,citecolor=NavyBlue,linkcolor=NavyBlue,pdfusetitle]{hyperref}
\usepackage[all]{hypcap}
\usepackage[inline]{enumitem}
\usepackage[utf8]{inputenc}
\usepackage{lipsum}
\usepackage{booktabs}
\usepackage{wasysym}

\usepackage{array}

\newcommand{\beq}{\begin{equation}}
\newcommand{\eeq}{\end{equation}}

\newcommand{\xobs}{x_\mathrm{obs}}
\newcommand{\xtrue}{{x}}
\newcommand{\sigmaobs}{\sigma_\mathrm{obs}}
\newcommand{\mupop}{\mu}
\newcommand{\sigmapop}{\sigma_\mathrm{pop}}
\newcommand{\sigmatot}{\sigma_\mathrm{tot}}
\newcommand{\mupopest}{\hat{\mupop}}
\newcommand{\sigmapopest}{\hat{\sigma}_\mathrm{pop}}
\newcommand{\sigmaprior}{\sigma_\mathrm{prior}}
\newcommand{\infd}{\,\mathrm{d}}
\newcommand{\priorw}{\Delta}
\newcommand{\qgr}{\mathcal{Q}_\mathrm{GR}}
\DeclareMathOperator{\var}{\mathrm{var}}

\newtoggle{commentsoff}
\togglefalse{commentsoff}
\ifdefined\nocomments
    \toggletrue{commentsoff}
\fi

\iftoggle{commentsoff}{
  \newcommand*{\mi}[1]{}
  \newcommand*{\wf}[1]{}
  \newcommand*{\kc}[1]{}
  \newcommand*{\comment}[1]{}
  
  \newcommand*{\todo}[1]{}
  \newcommand*{\warn}[1]{}

}{
  \newcommand*{\mi}[1]{{\color{magenta} [{\bf MAX}: #1]}}
  \newcommand*{\wf}[1]{{\color{RedOrange} [{\bf WILL}: #1]}}
  \newcommand*{\kc}[1]{\textcolor{ForestGreen}{[\textbf{KATERINA}: #1]}}
  
  \newcommand*{\comment}[1]{{\color{blue} [{\bf NOTE}: #1]}}
  \newcommand*{\warn}[1]{{\color{red} [{\bf WARNING}: #1]}}
  \newcommand*{\todo}[1]{{\color{red} [{\bf TODO}: #1]}}

}

\graphicspath{{./fig/}}

\newcommand{\dcc}{LIGO-P2200099}

\begin{document}

\title{Comparing Bayes factors and hierarchical inference for testing\\general relativity with gravitational waves}

\newcommand{\CCA}{\affiliation{Center for Computational Astrophysics, Flatiron Institute, 162 5th Ave, New York, NY 10010}}
\newcommand{\MIT}{\affiliation{LIGO Laboratory and Kavli Institute for Astrophysics and Space Research, Massachusetts Institute of Technology, Cambridge, Massachusetts 02139, USA}}
\newcommand{\STBR}{\affiliation{Department of Physics and Astronomy, Stony Brook University, Stony Brook NY 11794, USA}}
\newcommand{\CIT}{\affiliation{Department of Physics, California Institute of Technology, Pasadena, California 91125, USA}}
\newcommand{\CITLab}{\affiliation{LIGO Laboratory, California Institute of Technology, Pasadena, California 91125, USA}}

\author{Maximiliano Isi}
\email{misi@flatironinstitute.org}
\CCA

\author{Will M. Farr}
\email{will.farr@stonybrook.edu}
\CCA
\STBR

\author{Katerina Chatziioannou}
\email{kchatziioannou@caltech.edu}
\CIT
\CITLab

\hypersetup{pdfauthor={Farr, Isi, Chatziioannou}}

\date{\today}

\begin{abstract}
In the context of testing general relativity with gravitational waves,
constraints obtained with multiple events are typically combined either through
a hierarchical formalism or though a combined multiplicative Bayes factor. We
show that the well-known dependence of Bayes factors on the analysis priors in
regions of the parameter space without likelihood support can lead to strong
confidence in favor of incorrect conclusions when one employs the multiplicative
Bayes factor. Bayes factors ${\cal{O}}(1)$ are ambivalent as they depend
sensitively on the analysis priors, which are rarely set in a principled way;
additionally, combined Bayes factors $>{\cal{O}}(10^3)$ can be obtained in favor
of the incorrect conclusion depending on the analysis priors when many
$\mathcal{O}(1)$ Bayes factors are multiplied, and specifically when the priors
are much wider than the underlying population. The hierarchical analysis that
instead infers the ensemble distribution of the individual
beyond-general-relativity constraints does not suffer from this problem, and
generically converges to favor the correct conclusion. Rather than a naive
multiplication, a more reliable Bayes factor can be computed from the
hierarchical analysis.  We present a number of toy models showing that the
practice of multiplying Bayes Factors can lead to incorrect conclusions.
\end{abstract}

\maketitle

\section{Introduction}

With an increasing number of
LIGO-Virgo~\cite{TheLIGOScientific:2014jea,TheVirgo:2014hva} gravitational wave
(GW) observations, we can leverage the collective set of measurements to study
the properties of the astrophysical objects that generate
GWs~\cite{LIGOScientific:2020kqk,LIGOScientific:2021psn} and the validity of
general relativity
(GR)~\cite{Isi:2019asy,LIGOScientific:2020tif,LIGOScientific:2021sio}. Two broad
and complementary approaches exist for drawing inferences from sets of
detections. The first relies on the posterior distribution for some model and
its continuous parameters whose range of possible values encapsulates the
different physics we would like to study. The second phrases a question of
interest in the language of model selection between two discrete hypotheses to
compute Bayes factors (BFs). The latter approach is common in the context of
testing GR where one introduces a parametrized deviation of the signal as
predicted by
GR~\cite{LiEtAl:2012a,LiEtAl:2012b,AgathosEtAl:2014,Sampson:2013lpa,Sampson:2013jpa,PhysRevD.90.064009},
but further examples relate to higher-order modes of the
radiation~\cite{Chatziioannou:2019dsz}, GW memory~\cite{Hubner:2019sly}, the
neutron star equation of
state~\cite{DelPozzo:2013ala,Chatziioannou:2015uea,LIGOScientific:2019eut,Ghosh:2021eqv,Isi:2017equ,Callister:2017ocg},
gravitational lensing~\cite{Liu:2020par,Lo:2021nae}, the association between GWs
and potential electromagnetic counterparts~\cite{Ashton:2020kyr}, GW
ringdowns~\cite{Ota:2021ypb}, and the signal detection problem in
general~\cite{Veitch:2008wd,Veitch:2009hd,Cornish:2014kda,Isi:2018vst,Ashton:2019wvo,Pratten:2020ruz,Cornish:2020dwh}.

Although posteriors and BFs are mathematically related, in practice approaches
focusing on one or the other can come to seemingly different, and even
contradictory, conclusions. For example, Ref.~\cite{Bustillo:2020buq} considers
the case of testing the no-hair theorem with GW ringdowns and shows that BFs can
favor the incorrect conclusion even in cases where the posterior has minimal
support for it. The alternative of working directly with the posterior and
hierachical inference has been introduced in the context of tests of
GR~\cite{Isi:2019asy,LIGOScientific:2020tif,LIGOScientific:2021sio} after the
limitations of using BFs to combine information from multiple events were
pointed out in~\cite{Zimmerman:2019wzo}.  Specifically, multipyling BFs
corresponds to assuming that a GR deviation will manifest independently and
distributed according to the underlying prior in each observation
\cite{Zimmerman:2019wzo}.  In this paper, we further this line of argument to
highlight issues with the use of BFs in the context of nested models and show
that the hierarchical modeling of population distributions offers a more
flexible and reliable alternative.

BFs, or marginalized-likelihood ratios, provide a succinct way to compare the likelihood of two models in light of some data.
The BF comparing a hypothesis $\mathcal{H}_0$ to another $\mathcal{H}_1$ for some data $d$ is given by
\begin{equation}
\mathcal{B}^{0}_{1} \equiv \frac{P(d\mid \mathcal{H}_0)}{P(d\mid \mathcal{H}_1)} = \frac{\int p(d \mid \theta_0, \mathcal{H}_0)\, p(\theta_0|\mathcal{H}_0) \infd \theta_0}{\int p(d \mid \theta_1, \mathcal{H}_1)\, p(\theta_1|\mathcal{H}_1) \infd \theta_1}\, ,
\end{equation}
where the likelihoods are marginalized over some (potentially different) sets of parameters $\theta_{0/1}$ for the $\mathcal{H}_{0/1}$ hypotheses respectively.
The definition of the hypotheses encompasses the choice of parameter priors $p(\theta\mid \mathcal{H})$, as well as any other assumptions built into the functional form of the likelihoods $p(d\mid \theta, \mathcal{H})$.
A larger value of $\mathcal{B}^0_1$ (or, equivalently, its natural logarithm, $\ln \mathcal{B}^0_1$) indicates a preference for $\mathcal{H}_0$ over $\mathcal{H}_1$.
When enhanced with prior weights for each hypothesis, $P(\mathcal{H}_{0/1})$, this returns the betting odds in favor of one model over the other, conditional on the observed data---namely,
\begin{equation}
\mathcal{O}^{0}_{1} \equiv \frac{P(\mathcal{H}_0)P(d\mid \mathcal{H}_0)}{P(\mathcal{H}_1)P(d\mid \mathcal{H}_1)} = \frac{P(\mathcal{H}_0)}{P(\mathcal{H}_1)} \mathcal{B}^0_1\, .
\end{equation}
To avoid expressing an \emph{a priori} preference for either model, it is common to set $P(\mathcal{H}_0)=P(\mathcal{H}_1)$, and so $\mathcal{O}^0_1 = \mathcal{B}^0_1$.

BFs reduce the complicated problem of selecting between two models of reality to
a single number---a feature which lies at the core of its appeal, but also of
its shortcomings. Like other scalar statistics, interpreting this one number is
usually far from straightforward, leading to somewhat \emph{ad hoc} scales such
as~\cite{Kass:1995loi}. This difficulty is worsened by the fact that BFs
necessarily are affected by \emph{all} aspects of a model, including those
corresponding to less interesting regions of the parameter space like regions of
the prior space for which the likelihood offers no support. Consequently, BFs
can vary wildly with different choices of prior bounds, which are often set
arbitrarily. Since priors can rarely be set from first principles, calibrating
BFs tends to require large scale injection
campaigns~\cite{Isi:2018vst}---although this is only possible when the
injections themselves can be designed in a principled way (i.e., when we
actually know how to simulate expected astrophysical distributions). Even when
the model is specified correctly and we can take BFs at face value, the result
offers no insight as to why exactly one model is to be preferred.

All these drawbacks compound when one attempts to combine multiple observations
by multiplying BFs computed with a fixed prior, which enhances the sensitivity
to prior choices. Moreover, naive BF computations from collections of events
impose strong, generally unrealistic assumptions~\cite{Zimmerman:2019wzo}.
Multiplying BFs obtained from individual events results in a collective BF that
assumes the targeted effect (say, a deviation from GR) manifests independently
for each observation. On the other hand, generating a collective BF from the
product of likelihoods from multiple measurements presumes that the effect
manifests identically for all observations. Neither of these are valid
assumptions in general~\cite{Zimmerman:2019wzo}.  In general the degree to which
a targeted effect appears independently versus identically in each observation
is something that should be leared from the data; this insight leads directly to
hierarchical modeling \cite{James:1961,Lindley:1972,Efron:1977,Rubin:1981}.

Rather than assuming a fixed and known distribution (e.g., all events are the
same, or all the events are different), a hierarchical analysis works by
inferring the underlying distribution of the parameter whose values encode the
targeted hypotheses. For example, in the context of tests of GR, a parameter $x$
may represent the magnitude of a GR violation, so that the GR (non-GR)
hypothesis implies $x=0$ ($x\neq0$); likewise, when searching for GW memory,
this could be the amplitude of the effect in question. Once this parameter is
identified, we can use hierarchical inference to characterize its values across
the observed events---using the collection of measurements holistically to infer
the distribution of $x$. The challenge here lies in choosing a suitable
parametrization for this underlying distribution.

We can circumvent this through a moment expansion: as a first approximation, we
will only be interested in recovering the mean and standard deviation of the
unknown distribution, so we can parametrize it as a Gaussian, whose mean and
standard deviation ($\mupop$, $\sigmapop$) we are to measure from the
data~\cite{Isi:2019asy}; the null hypothesis will often be constructed so that
it is recovered for $\mu=\sigmapop=0$. This approach allows us to study the
population of measurements without assuming the targeted effect manifests either
identically or distinctly for all events, learning more about the population
distribution, which now plays the role of a prior, with each new observation.

In this paper, we compare the BF and hierarchical approaches directly in the context of multiple GW observations, and argue that BFs are unreliable in any context in which the prior does not adapt to the observations at hand.
We show that in such context, BFs do not have the right scaling with the number of events, even in simple situations, due to their inherently strong dependence on the (fixed) priors in regions of no likelihood support.
We show that hierarchical posteriors do not suffer from such limitations as they rely on ``priors'' that are inferred from the data, and have a weaker dependence on hyperparameter assumptions.

We begin in Sec.~\ref{sec:single} by reviewing some basic properties of BFs obtained from single observations and their interplay with Occam penalties.
Then, in Sec.~\ref{sec:toy-model}, we study the scaling of BFs when combining multiple observations under two example distributions for the true parameters under consideration; we show that this approach can often lead to the incorrect conclusion.
We also consider the same scenarios under the hierarchical approach, showing that it does not suffer from the same drawbacks.
In Sec.~\ref{sec:fencepost}, we consider a final model that cannot be obtained as a special case of the distribution assumed by the hierarchical analysis, showing that the hierarchical approach yields the correct result even then.
We conclude in Sec.~\ref{sec:conclusion}.

\section{Single event: Occam penalty vs goodness-of-fit}
\label{sec:single}
\begin{figure}
  \centering
  \includegraphics[width=0.5\textwidth]{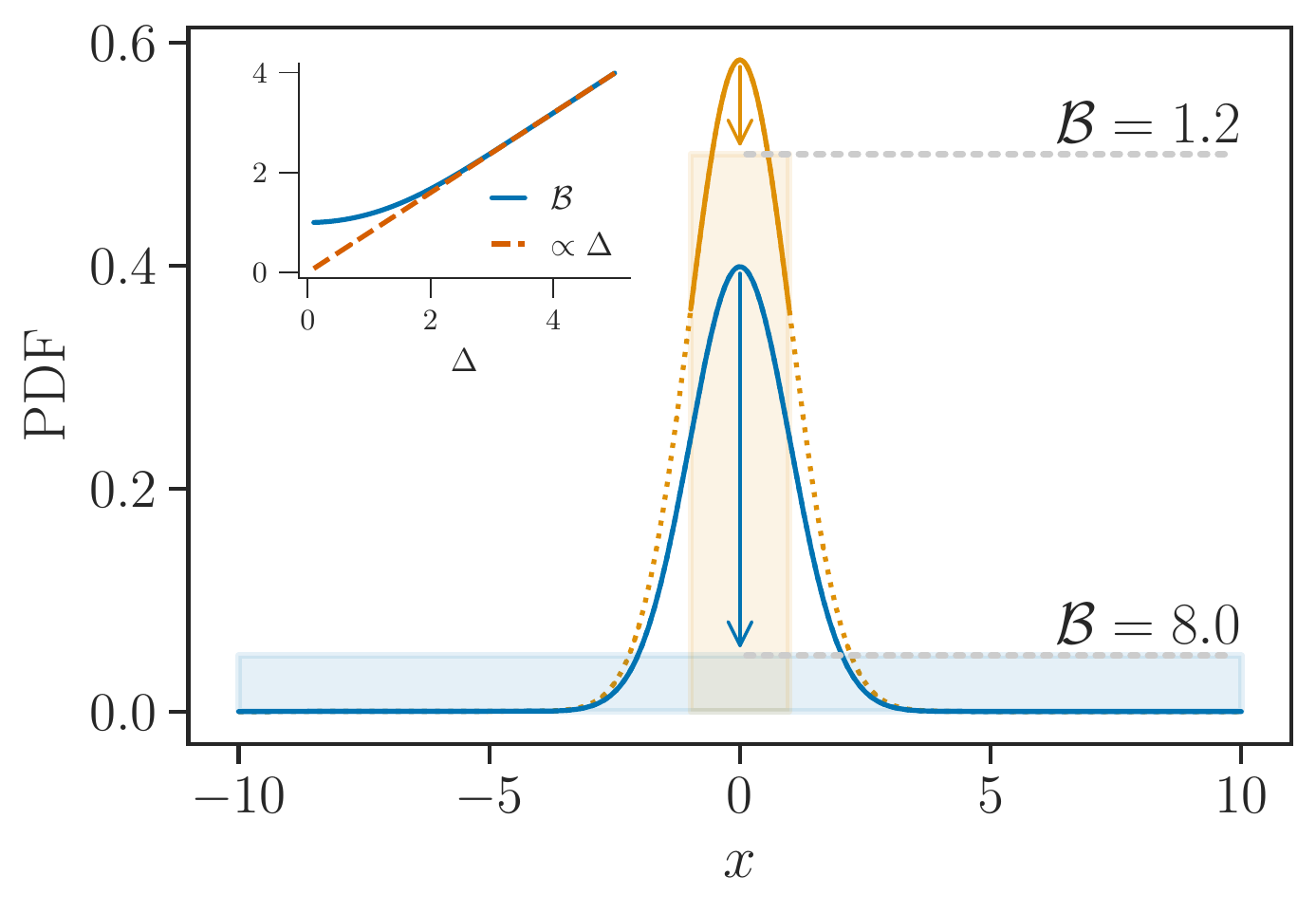}
  \caption{Bayes factor computation as a Savage-Dickey density ratio. The BF ($\mathcal{B}$) comparing a model in which some parameter $x$ takes the value $x = 0$ to one in which $x\neq 0$ is  given by the ratio of the posterior to the prior evaluated at $x = 0$. The main panel shows the case of a normal posterior truncated by a broad prior (half-width $\Delta = 10$, in blue) and a narrow prior ($\Delta = 1$, in orange). The inset shows the scaling of $\mathcal{B}$ as a function of $\Delta$ (solid curve); for broad priors, $\mathcal{B} \propto \Delta$ (dashed line). In this toy example, we assume the posterior peaks always at $x=0$.
  }
  \label{fig:eyecandy}
\end{figure}

Before tackling the problem of multiple observations, we first review the behavior of BFs computed from a single event.
A typical situation that is simple to understand analytically is that of nested models, i.e., two models constructed so that one can be recovered as a special case of the other.
Parametrized tests of GR, for example, typically involve nested hypotheses wherein the non-GR model is characterized by all the usual GR signal parameters plus one or more additional variables $x_{\rm nGR}$ that quantify the deviation from GR~\cite{Blanchet:1994ez,Blanchet:1994ex,Arun:2006hn,Arun:2006yw,Yunes:2009ke,Mishra:2010tp,Li:2011cg,Li:2011vx,Agathos:2013upa}.
These parametrizations are usually constructed so that GR is recovered when the deviation parameters vanish.
Then, to determine whether GR is favored, the data are analyzed with some broad (typically flat) prior on the deviations in order to compute BFs comparing $x_{\rm nGR} =0$ to $x_{\rm nGR} \neq 0$.

In that spirit, consider some real-valued parameter $x$ and two related hypotheses $\mathcal{H}_{x=0}$ and $\mathcal{H}_{x\neq0}$, respectively defined to imply $x=0$ and $x\neq0$, with some prior over $x$ and any other relevant parameters.
Since $x=0$ is a special case of the model in which $x$ is allowed to vary over a broad range including the origin, we say the hypotheses are nested.%
\footnote{Here we are identifying $\mathcal{H}_{x\neq0}$ with the model in which $x$ is allowed to vary freely; this is legitimate because the point $x=0$ is a set of measure zero, so it does not need to be explicitly excised from the arbitrary-$x$ model.}
In that case, the BF comparing the two is given exactly by the Savage-Dickey ratio~\cite{Dickey:1971,Verdinelli:1995},
\begin{equation} \label{eq:sd}
\mathcal{B}^{x=0}_{x\neq 0} = \frac{p(x=0 \mid d)}{p(x=0)}\, ,
\end{equation}
where $p(x=0 \mid d)$ is the marginal posterior and $p(x=0 )$ is the prior, both evaluated at $x=0$.
In other words, the BF in favor of $x=0$ is simply the ratio of the posterior to the prior evaluated at the origin.

We can elucidate the role of the prior in Eq.~\eqref{eq:sd} by considering a specific functional form.
For simplicity, assume the marginal posterior is given by a standard normal distribution, truncated symmetrically around the origin by some uniform prior of half-width $\priorw$ (i.e., flat in $-\priorw < x < \priorw$).
Then, the BF can be computed analytically to yield
\begin{equation} \label{eq:bf_simple}
\mathcal{B}^{x=0}_{x\neq 0} = \frac{1}{\sqrt{2\pi}}\frac{2\priorw}{\Phi(\priorw) - \Phi(-\priorw)} = \sqrt{\frac{2}{\pi}} \frac{\priorw}{\mathrm{erf}(\priorw/\sqrt{2})}\, ,
\end{equation}
in terms of the standard cumulative distribution function $\Phi(x) \equiv \left(1 + \mathrm{erf}(x/\sqrt{2})\right)/2$ and the error function $\mathrm{erf}(x) \equiv (2/\sqrt{\pi}) \int_0^x \exp(-y^2) \infd y$.
Since $\mathrm{erf}(\priorw/\sqrt{2}) \to 1$ for large $\priorw$, the BF can be made to favor $x = 0$ with arbitrarily-high confidence by sufficiently broadening the prior---in fact, $\mathcal{B}^{x=0}_{x\neq 0} \propto \priorw$ in the large $\Delta$ limit.
We illustrate this in Fig.~\ref{fig:eyecandy}.

The dependence on the prior range is a general feature not specific to our
example: the same data can produce arbitrary odds in favor of a specific value
of a parameter ($x=0$ here) relative to a model with increasing prior volume
(proxied by $\priorw$). This is related to the concept of the Occam penalty in
Bayesian inference: BFs do not only favor the model that fits the data best, but
also the one that is simplest---where simplicity is defined as a model's ability
to fit the data without having to significantly constrain its parameters
relative to their a priori allowed values. The interplay between goodness of fit
and Occam penalty creates the possibility of a BF that strongly favors the
incorrect conclusion. In the context of testing GR, if the theory is indeed
violated, then some observation can give $p(x=0\mid d) \ll 1$ in
Eq.~\eqref{eq:sd}; yet, this can always be countered with a wide enough prior
that makes the Occam penalty $1/p(x=0)$ so large that goodness of fit cannot
overcome it.  This is the expected manifestation of the Occam penalty in BFs;
the failure is symptomatic of a mismatch between the implemented prior and the
observer's expectation.

Since we generally have little a priori information about the nature of possible
beyond-GR effects, and these effects are not strongly constrained by the
likelihood of an individual measurement, a natural inclination is to make the
prior much wider than the likelihood. However, the argument above shows that a
broad prior prevents us from detecting small deviations from GR, which is an
important regime for tests of GR~\cite{Perkins:2022fhr}---either because GR is
close to correct or because of selection effects that disfavor the detection of
signals with morphology far from GR.  The same tension arises in other contexts
where BFs are used without a principled prior. In the next section, we show that
this behavior is not unique to single-event analyses but carries over to
combined constraints.

\section{Combining events under a broad prior}
\label{sec:toy-model}

We now turn to collections of measurements and show that the ``combined''
multiplicative BF does not have the correct scaling in the regime of interest,
with support accumulating in favor of the null hypothesis even when this
conclusion is incorrect. Combining BFs from multiple events will only lead to
the right conclusion when the deviation (e.g., the deviation from GR) is large
enough to be apparent in individual posteriors, negating the need for combining
observations in the first place. We then show that the hierarchical approach is
not susceptible to this issue.

Again, consider a single parameter, $\xtrue$, that stands in for the magnitude of the GR deviation or any other effect of interest, and let $\xtrue = 0$ be our null hypothesis (e.g., GR is correct).
We conduct experiments to measure $\xtrue$, and assume additive Gaussian noise so that the observed value $\xobs$ is normally distributed
about the true value $\xtrue$ with standard deviation (measurement noise)
$\sigmaobs$:
\begin{equation}
  \label{eq:toy-likelihood}
  p\left( \xobs \mid \xtrue \right) = \frac{1}{\sqrt{2\pi \sigmaobs^2}} \exp\left( - \frac{\left( \xobs - \xtrue \right)^2}{2 \sigmaobs^2} \right).
\end{equation}
In the previous section, we considered a simplified case of this model, in which we had a single measurement with $\xobs=0$ and $\sigmaobs=1$.

We consider two ``populations'' for the true values of $\xtrue$ under repeated measurements, i.e., the true magnitude of the GR deviation for each observed GW event.
In the first the true value of $\xtrue$ is fixed to some value $\xtrue_0$ for each measurement so that the population distribution is a Dirac delta,
\begin{equation} \label{eq:munotzero}
p_1\left( \xtrue \right) = \delta\left( \xtrue - \xtrue_0 \right).
\end{equation}
In the second, the value of $\xtrue$ is randomly distributed with mean zero and
standard deviation $\sigma_0$ for each measurement, so that the population
distribution is
\begin{equation} \label{eq:sigmanotzero}
  p_2\left( \xtrue \right) = \frac{1}{\sqrt{2\pi \sigma_0^2}} \exp\left( - \frac{\xtrue^2}{2 \sigma_0^2} \right).
\end{equation}
In the language of tests of gravity, the two models recover GR (our null hypothesis) when $\xtrue_0 = 0$ or $\sigma_0 = 0$, respectively.
In the second model, the \emph{mean} deviation from GR vanishes, but the actual variation in a given measurement fluctuates.  In both cases the deviation parameter $\xtrue$ is
assumed independent and identically distributed (iid) for each measurement, so there
is some prior choice such that multiplying BFs for repeated
measurements is optimal~\cite{Zimmerman:2019wzo}.\footnote{The optimal prior is
the actual population from which the true parameters controlling the
measurements are iid draws.}
However, that prior is unknown in realistic situations because the true distribution of GR deviations is not known a priori.
Rather, in all cases we assume a measurement is analyzed with a flat prior on $-\priorw < \xtrue < \priorw$, following common practice.

\subsection{Bayes factors}

With that prior, the BF between the null and generalized hypotheses (for concreteness, ``GR'' and ``non-GR'') for a given
observation $\xobs$ is a generalization of Eq.~\eqref{eq:bf_simple}
\begin{multline}
  \label{eq:sd-bf-toy}
  \mathcal{B}^\mathrm{GR}_\mathrm{nGR} = \frac{2 \priorw}{\sqrt{2\pi \sigmaobs^2} \left( \Phi\left( \frac{\priorw - \xobs}{\sigmaobs} \right) - \Phi\left( \frac{-\priorw - \xobs}{\sigmaobs} \right) \right)} \\ \times \exp\left( - \frac{\xobs^2}{2\sigmaobs^2} \right).
\end{multline}
When the prior is very broad, $\priorw \gg \xobs$, and the Gaussian posterior on $\xtrue$ is not meaningfully truncated by the prior, then the BF simplifies to
\begin{equation}
  \label{eq:bf-approx-toy}
  \mathcal{B}^\mathrm{GR}_\mathrm{nGR} \simeq \frac{2 \priorw}{\sqrt{2 \pi \sigmaobs^2}} \exp\left( - \frac{\xobs^2}{2\sigmaobs^2} \right),
\end{equation}
corresponding to the $\mathcal{B}^{x=0}_{x\neq0} \propto \priorw$ limit discussed in the previous section.
Ensuring $\priorw \gg \xobs$ is a common analysis choice because this prior
permits the true value of $\xtrue$ to correspond to the observed value
$\xobs$ which is the value of $\xtrue$ that maximizes the likelihood in each
observation.
A broad prior is also desired when combining multiple observations in order to accommodate the expected scatter in the individual likelihoods.

If we choose to combine observations by adding log BFs\footnote{In this paper,
all logarithms are natural logarithms.} (equivalently, multiplying BFs), it is
sufficient to compute the expected value of an individual log BF under the true
deviations; the expected total log BF will then be the expected individual log
BF times the number of events. The expected value of the log of Eq.\
\eqref{eq:sd-bf-toy} is not expressible in closed form, but can be
straightforwardly computed numerically. In the limit that $\priorw \gg \xobs$,
the expected log BFs under our two populations of GR deviations become
\begin{align}
  \label{eq:approx-ex-bf-1}
  \left. \left \langle \ln \mathcal{B}^\mathrm{GR}_\mathrm{nGR} \right\rangle \right|_{\xtrue_0} &\simeq \ln \frac{2 \priorw}{\sqrt{2\pi \sigmaobs^2}} - \frac{\sigmaobs^2 + \xtrue_0^2}{2 \sigmaobs^2} \nonumber\\
  &\simeq \ln \frac{\priorw}{\sigmaobs} - 0.23 - \frac{\sigmaobs^2+\xtrue_0^2}{2 \sigmaobs^2},
\end{align}
and
\begin{align}
  \label{eq:approx-ex-bf-2}
  \left. \left \langle \ln \mathcal{B}^\mathrm{GR}_\mathrm{nGR} \right\rangle \right|_{\sigma_0} &\simeq \ln \frac{2 \priorw}{\sqrt{2\pi \sigmaobs^2}} - \frac{\sigmaobs^2 + \sigma_0^2}{2\sigmaobs^2} \nonumber \\
  &\simeq \ln \frac{\priorw}{\sigmaobs} - 0.23 - \frac{\sigmaobs^2+\sigma_0^2}{2\sigmaobs^2}.
\end{align}
From the above, we can see that, whenever the GR deviation is nonzero but small
enough to be undetectable in a single observation ($0 < \xtrue_0, \sigma_0 \ll
\sigmaobs$), then for $\priorw \gtrsim 2 \sigmaobs$ the expected log BF
is positive, and evidence accumulates \emph{in favor} of GR even though there is
a deviation.  For deviations that are marginally detectable in a single
observation $\xtrue_0, \sigma_0 \simeq \sigmaobs$, choosing a wide,
uninformative prior $\priorw \gtrsim 3.5 \sigmaobs$ (which, recall, is necessary
to ensure that all observations $\xobs$ in a modest-sized catalog of
observations are within the prior range) will result in evidence that
accumulates \emph{against} modifications to GR.

\begin{figure*}
  \includegraphics[width=\columnwidth]{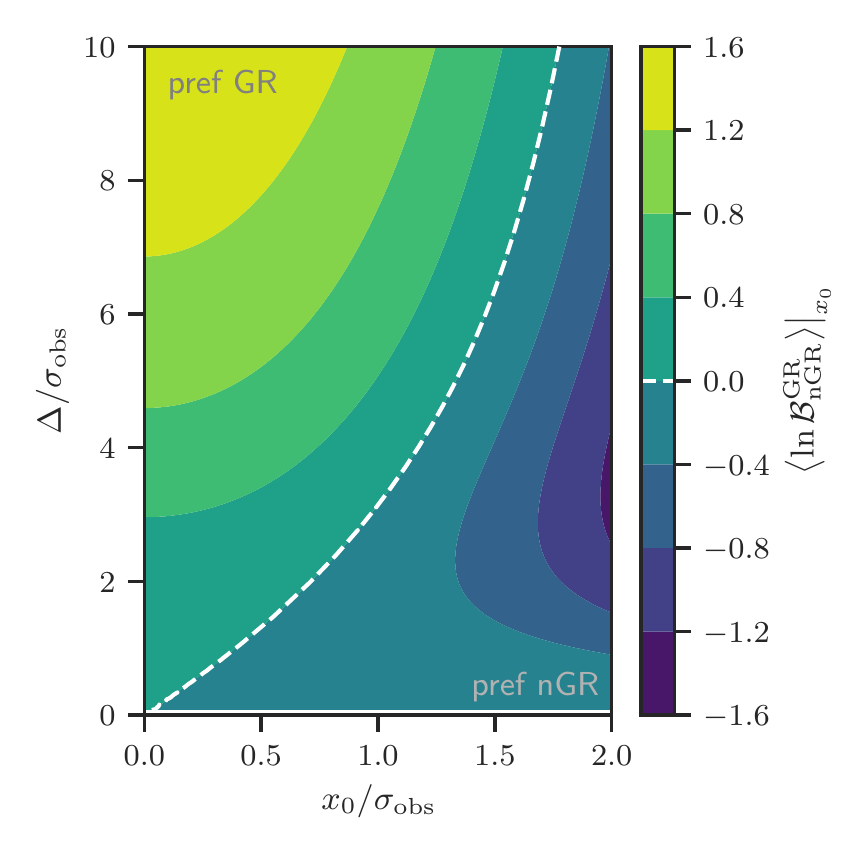}
  \includegraphics[width=\columnwidth]{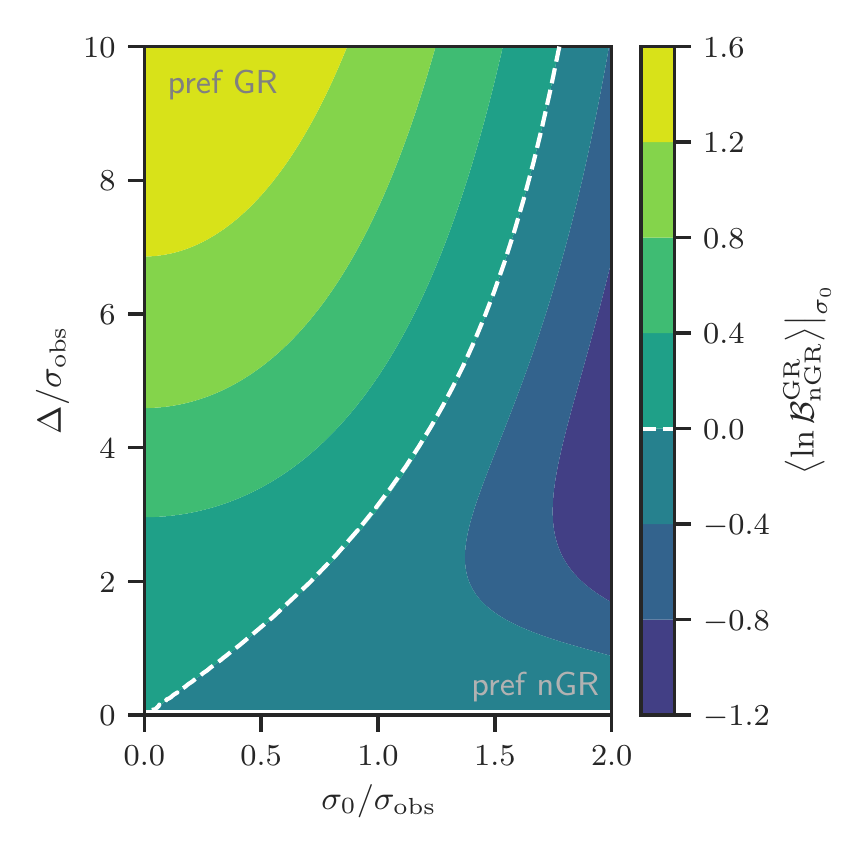}
  \caption{\label{fig:toy-BF-parameter-scan} The expected log BF in
  favor of GR from the toy model discussed in Sec.~\ref{sec:toy-model} (color) as a function of the prior width $\priorw$ (ordinate) and the true
  deviation parameter $\xtrue_0$ (abscissa left) or the scatter in the zero-mean
  deviation parameter $\sigma_0$ (abscissa right). Both quantities are shown relative to
  the measurement uncertainty on the deviation parameter $\sigmaobs$.  For any
  non-zero value of the deviation parameter or its scatter, there is a prior
  width that results in incorrectly accumulating evidence \emph{for} GR (above dashed line); moreover, for reasonable prior widths of a few times the
  measurement uncertainty (so that observed values of the deviation parameter
  lie within the prior range and the prior is therefore broad and uninformative)
  the deviation parameter must be large enough to be marginally detectable in a
  single measurement ($\xtrue_0, \sigma_0 \sim \sigmaobs$) before evidence
  against the incorrect GR model accumulates on average over many stacked
  measurements. Fig.~\ref{fig:toy-BF-slice} demonstrates this effect for $\priorw = 5 \sigmaobs$.}
\end{figure*}

An exact calculation of the expected log BF without the assumption
that $\priorw \gg \xobs$ appears in Fig.~\ref{fig:toy-BF-parameter-scan}.
Interestingly, any value of $\priorw$ will
accumulate evidence for GR when $\xtrue_0 = 0$ or $\sigma_0 = 0$; but if
$\xtrue_0 > 0$ or $\sigma_0 > 0$, so that the GR model is incorrect, prior
choices that encompass deviation parameter values comparable to those observed,
i.e., $\priorw \simeq \mathrm{few} \times \sigmaobs$, accumulate evidence for the
incorrect GR model unless the deviation parameter is comparable to or larger
than the observational uncertainty.  In this regime, either the BF
fails to select the correct theory (non-GR) on average (with more and more
certainty as the number of observations grows); or the deviation is so large
that it is marginally detectable (i.e., ``${\sim} 1\sigma$'') with a single
observation.  Multipliying BFs in this situation is counter-productive,
and the best constraint is achieved with a single measurement.

\begin{figure*}
  \includegraphics[width=0.9\columnwidth]{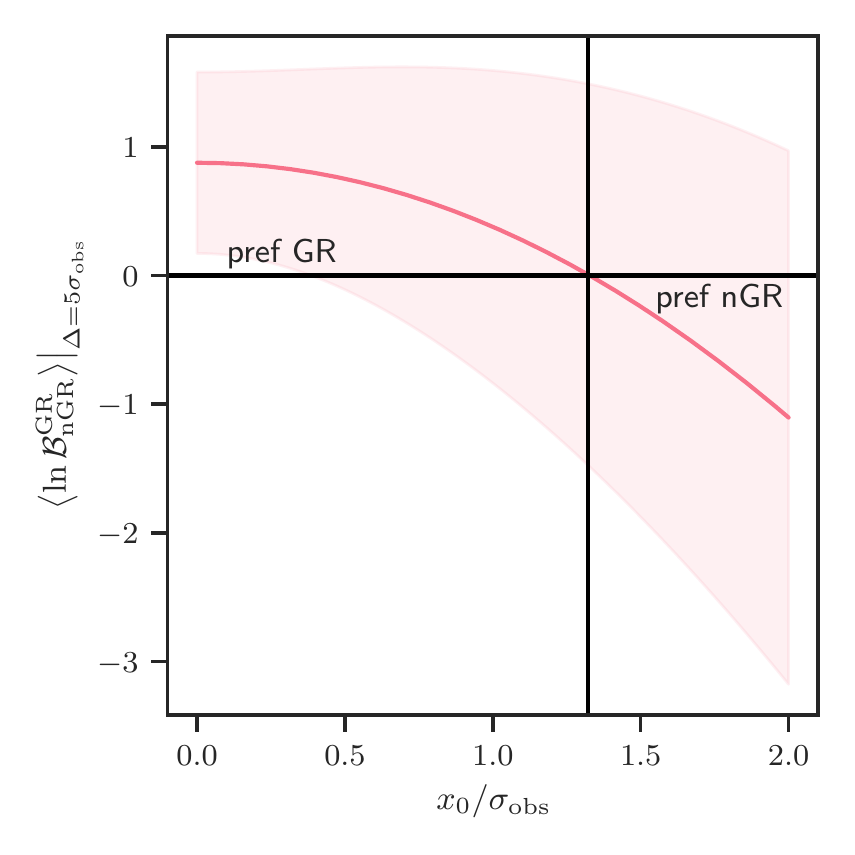}\qquad
  \includegraphics[width=0.9\columnwidth]{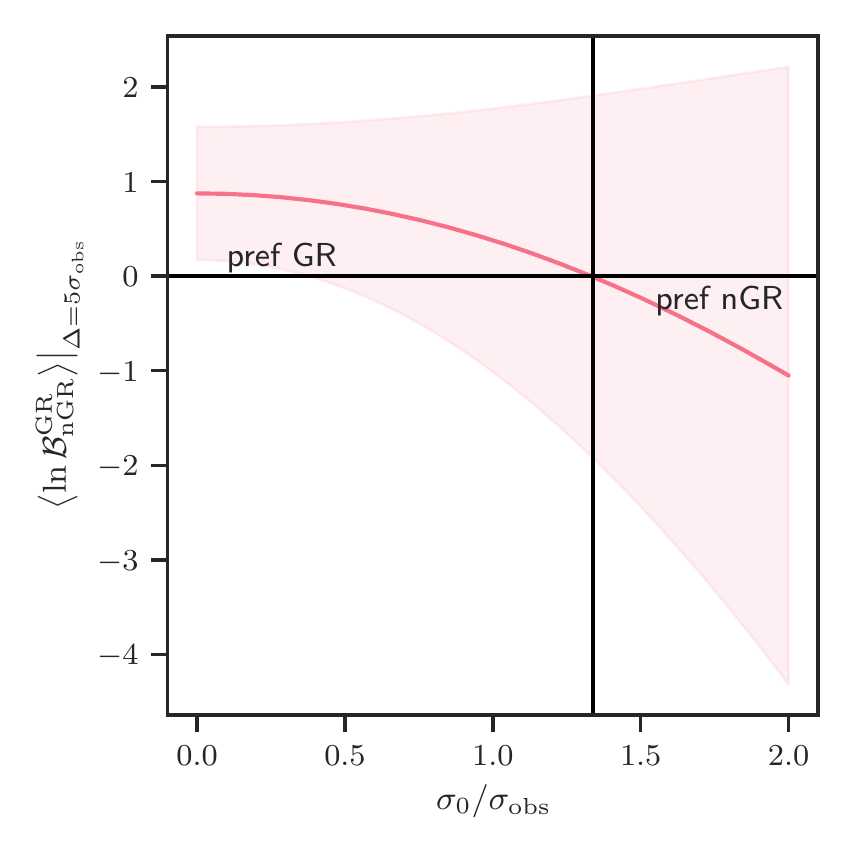}
  \caption{\label{fig:toy-BF-slice} The expected log BF for the GR
  model when $\priorw = 5 \sigmaobs$ (a broad, uninformative prior) versus the
  fixed value of the deviation parameter ($\xtrue_0$, left) or scatter in the
  zero-mean deviation parameters ($\sigma_0$, right).  For this choice of prior
  width, the average log BF favors the incorrect GR model until the
  deviation parameter is $\xtrue_0, \sigma_0 \gtrsim 1.3 \sigmaobs$, marginally
  detectable in a single measurement. Shading shows the expected variation in the log BF ($\pm1\sigma$).}
\end{figure*}

\begin{figure}
  \centering
          \includegraphics[width=0.95\columnwidth]{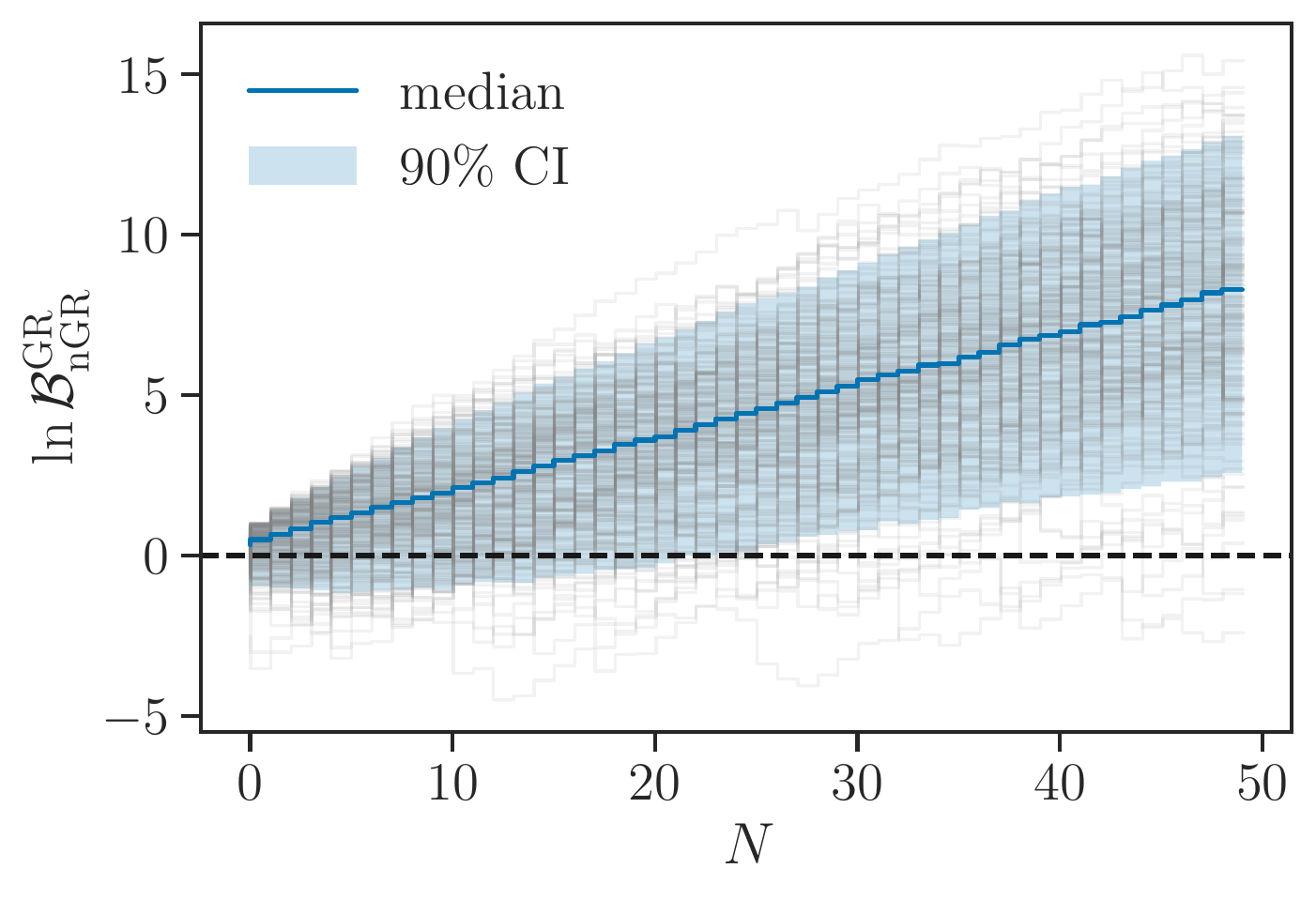}
  \caption{Log BF for a growing catalog of $N$ observations in the case where the null hypothesis (GR) is correct, and where $\Delta = 2\sigmaobs$, for 200 simulations (represented by each thin gray trace). On average, the log BF favors the right hypothesis (blue line), but there is enough variance in this quantity that some of the individual catalogs favor the wrong hypothesis, even for a moderately high number of detections (traces below the dashed line).}
  \label{fig:bf_spread}
\end{figure}

Even when the log BF is expected to take the correct sign on average, the result will vary for any given set of detections.
In the regime where $\Delta \gg x_\mathrm{obs}$, we can quantify this through the variance associated with the means in Eqs.~\eqref{eq:approx-ex-bf-1} and \eqref{eq:approx-ex-bf-2}.
For the former, this is just
\begin{equation}
\left.\mathrm{var}\left(\ln \mathcal{B}^{\rm GR}_{\rm nGR}\right) \right|_{\xtrue_0} \simeq \left(\frac{\xtrue_0}{\sigmaobs}\right)^2 + \frac{1}{2}\, ,
\end{equation}
while for the latter we have
\begin{equation}
\left.\mathrm{var}\left(\ln \mathcal{B}^{\rm GR}_{\rm nGR}\right) \right|_{\sigma_0} \simeq \frac{\left(\sigma_0^2 + \sigmaobs^2\right)^2}{2\sigmaobs^4}\, .
\end{equation}
These scalings are illustrated in Fig.~\ref{fig:toy-BF-slice} for $\Delta = 5 \sigmaobs$.
On the left hand side, the variance asymptotes to $1/2$ for decreasing $\sigmaobs$ and fixed intrinsic scatter $\sigma_0$; on the right hand side, the BF variance keeps increasing as we make the individual measurements more precise (decreasing $\sigmaobs$) because for narrower posteriors the value at $x=0$ varies more drastically from event to event.
In the case where GR is correct ($\xtrue_0=\sigma_0=0$), the scatter in the single-event log BF is large enough that it is not extremely uncommon for moderately large catalogs to yield evidence for the wrong hypothesis, even when the null hypothesis is correct (Fig.~\ref{fig:bf_spread}).

The qualitative behavior here can be understood in the context of
\cite{Zimmerman:2019wzo}.  The flat prior on $x$ implicitly assumes that
each observation has a true $x$ parameter that is independent of other
observations and uniformly distributed on $\left( -\priorw, \priorw \right)$.  If
one chooses $\priorw$ large enough to include all event likelihoods and not truncate them, then the flat prior for
the true parameter implicitly demands that most of the true deviation parameters
are comparable to $\priorw$.  If, instead, the deviation parameters are
considerably smaller than the observational uncertainty $\sigmaobs$---which is
the regime where stacking multiple events \emph{should} be the most
beneficial---then the assumption is so badly violated that pooling observations
prefers the incorrect model where the deviation parameters are zero to the
even-less-correct model where the deviation parameters are iid uniform on
$\left( -\priorw, \priorw \right)$.

The solution to this problem is to allow the assumed population to adapt its
properties to the stacked set of observations, for example by allowing its
location and scale to fit the set, as suggested in~\cite{Isi:2019asy}.  This
effectively constructs a model that better represents our beliefs about the
combined data set.

\subsection{Hierarchical treatment}

We now revisit the two experiments above using a hierarchical model for the distribution of GR deviations, instead of computing BFs with a uniform prior.
Assuming we are only interested in the first two moments of the distribution, we parametrize the true deviations as drawn from a Gaussian such that $\xtrue \sim \mathcal{N}(\mu, \sigmapop)$~\cite{Isi:2019asy}.
The goal will be to infer the $\mupop$ and $\sigmapop$ hyperparameters from the collection of measurements, and to quantify agreement with the null hypothesis $\mu=\sigmapop=0$ based on the corresponding 2D posterior.
The two non-GR models we considered above are encompassed within this parametrization when $(\mupop=\xtrue_0,\, \sigmapop=0)$ and $(\mupop=0,\, \sigmapop=\sigma_0)$.\footnote{The next section examines a non-GR model that is not fully encompassed in the Gaussian hierarchical model.}

As before, we assume that the observed value $\xobs$ in an individual event is normally distributed around the true value per Eq.~\eqref{eq:toy-likelihood}, and we take the true values themselves to be distributed normally given $\mupop$ and $\sigmapop$. 
In that case, the likelihood for a given observation becomes (see Appendix~\ref{app:hier})
\begin{equation} \label{eq:hier-like}
p(\xobs \mid \mupop, \sigmapop, \sigmaobs) = \frac{1}{\sqrt{2\pi \sigmatot^2}} \exp \left(-\frac{\left(\xobs - \mupop\right)^2}{2\sigmatot^2}\right) ,
\end{equation}
where $\sigmatot^2 \equiv \sigmaobs^2 + \sigmapop^2$ is the total variance arising from the combination of statistical uncertainty and the intrinsic population scatter.
Unlike in Eq.~\eqref{eq:toy-likelihood}, the true value $\xtrue$ for the individual measurement does not appear in this likelihood because we have marginalized over it and replaced it with the hyperparameters $\mupop$ and $\sigmapop$.

\begin{figure*}
  \includegraphics[width=\columnwidth]{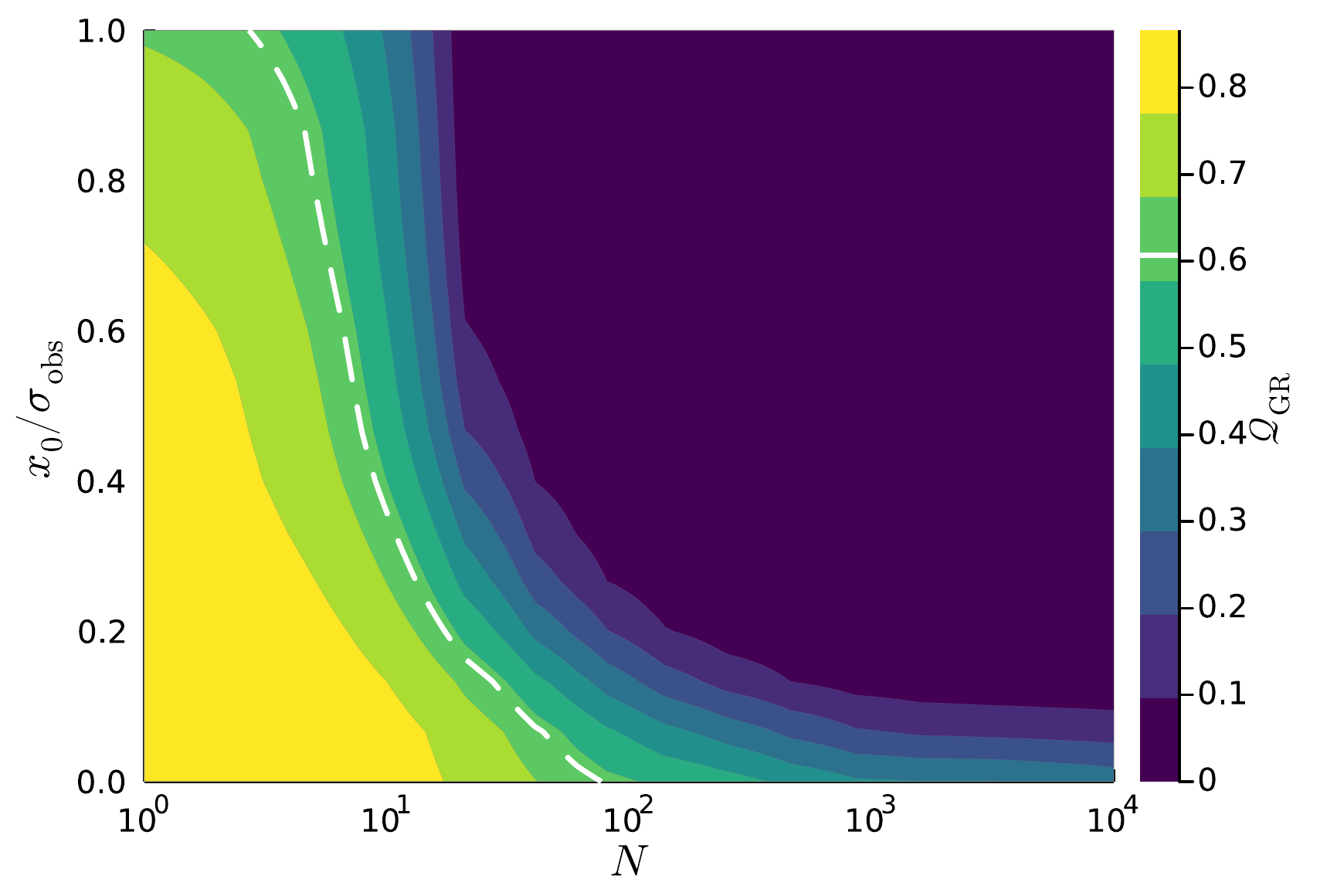}
  \includegraphics[width=\columnwidth]{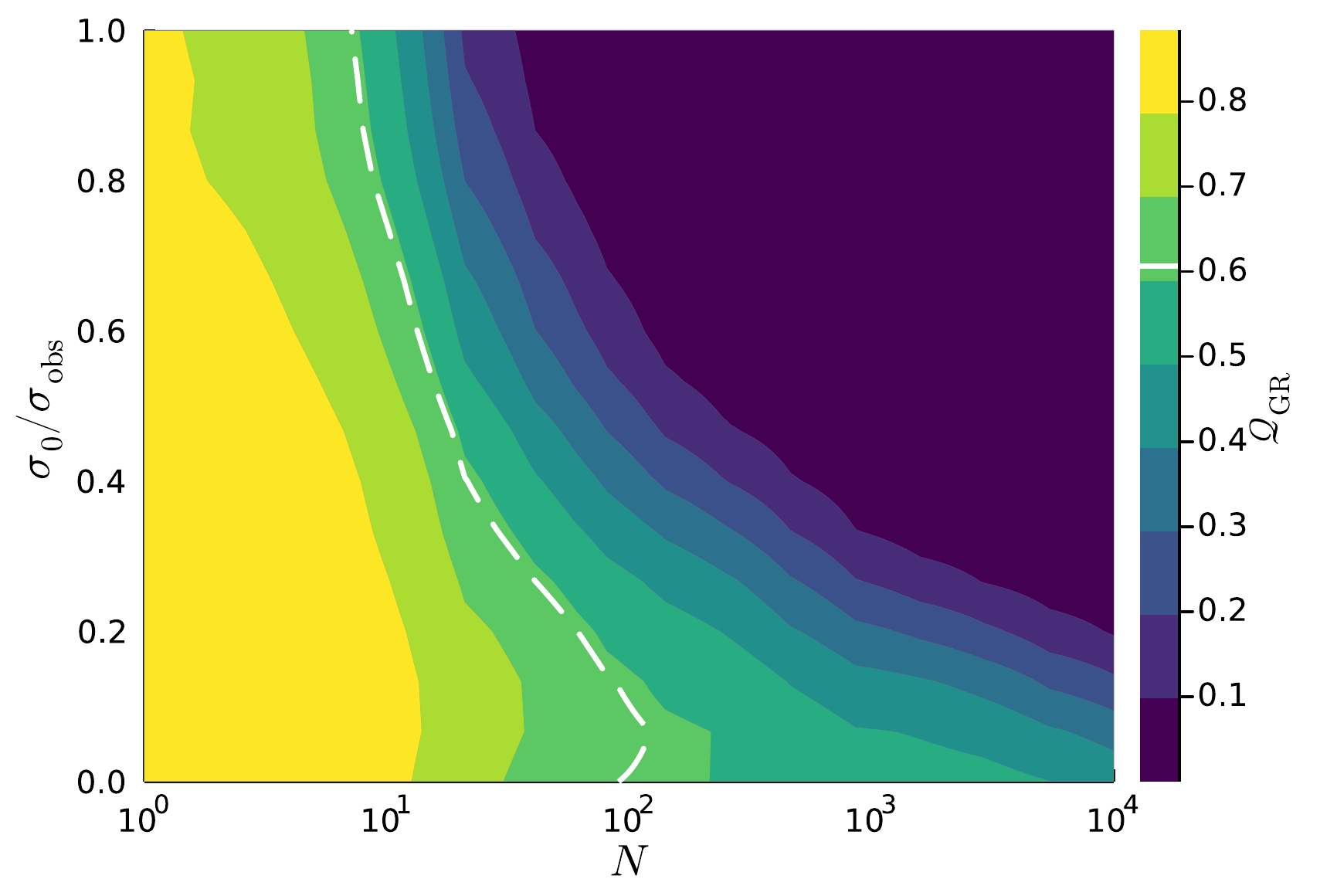}
  \caption{\label{fig:toy-qgr} The GR quantile $\qgr$ (color) obtained in a hierarchical analysis of data in which there is a fixed deviation for all events ($\xtrue_0$, ordinate left) or a scatter across events ($\sigma_0$, ordinate right), as a function of the number of events ($N$, abscissa).
  The reported value of $\qgr$ is the median over 50 simulated catalogs with $N$-events.
  Low values of $\qgr$ indicate that the null hypothesis ($\mupop=\sigmapop=0$) is disfavored; for reference, the dashed line marks the point at which GR is disfavored at $1\sigma$, i.e., $\qgr = \exp(-1/2) \approx 0.61$.
  Unlike in Fig.~\ref{fig:toy-BF-parameter-scan}, given enough $N$ we always detect the deviation, even when $\xtrue_0,\sigma_0 < \sigmaobs$.
  }
\end{figure*}

Consider the same two true distributions for $x$ above, which depart from GR as given by Eqs.~\eqref{eq:munotzero} and \eqref{eq:sigmanotzero}.
We simulate catalogs of detections following these distributions and analyze them hierarchically with the likelihood of Eq.~\eqref{eq:hier-like}, and Gaussian priors on $\mupop$ and $\sigmapop$ (with zero mean and standard deviation equal to $\sigmaobs$, restricting to positive values for $\sigmapop$).
Although other choices are possible, it is natural to tie the scale of the hyperprior to the measurement uncertainty because this is the only scale built into the problem a priori.
Furthermore, this choice is guaranteed to be sufficiently broad in the small-deviation regime ($\xtrue_0,\sigma_0 < \sigmaobs$) in which we are interested, and smooth enough to accommodate larger deviations if needed.

We quantify agreement with the null hypothesis through the marginal posteriors for $\mupop$ and $\sigmapop$, as well as the credible level at which GR is recovered in the 2D posterior for those two quantities (the 2D quantile, $\qgr$ in~\cite{LIGOScientific:2020tif}), defined as
\begin{equation}
\qgr \equiv \int_{p < p(0, 0)} p(\mupop, \sigmapop \mid \xobs, \sigmaobs) \infd \mupop \infd \sigmapop \, ,
\end{equation}
where the shorthand ``$p < p(0, 0)$'' stands in for values of $\mu$ and $\sigmapop$ such that $p(\mupop, \sigmapop \mid \xobs, \sigmaobs) < p(\mupop=0, \sigmapop = 0 \mid \xobs, \sigmaobs)$.
Given this definition, a value of $\qgr = 1$ means that the posterior peaks at the origin $\mupop=\sigmapop=0$, while $\qgr = 0$ means that the posterior offers no support for that point (i.e., a higher value of $\qgr$ implies better agreement with GR).
In all cases, we simulate each catalog of $N$ observations 50 times and report medians over the ensemble.

Unlike with BFs, there are no values of $\xtrue_0$ or $\sigma_0$ for which the hierarchical analysis converges to the wrong answer given enough observations.
This is apparent from Fig.~\ref{fig:toy-qgr}, which shows the value of $\qgr$ as a function of the deviation magnitude ($\xtrue_0/\sigmaobs$ or $\sigma_0/\sigmaobs$) and the number of observations: even for deviations small relative to the individual-measurement uncertainty, $\qgr$ approaches zero for large $N$---indicating that the posterior offers little support for $\mupop = \sigmapop=0$, in tension with GR.
For small catalogs ($N \leq 10$), the value of $\qgr$ is more strongly influenced by the prior, to a greater or lesser extent depending on the magnitude of the deviation.

\begin{figure*}
  \includegraphics[width=\columnwidth]{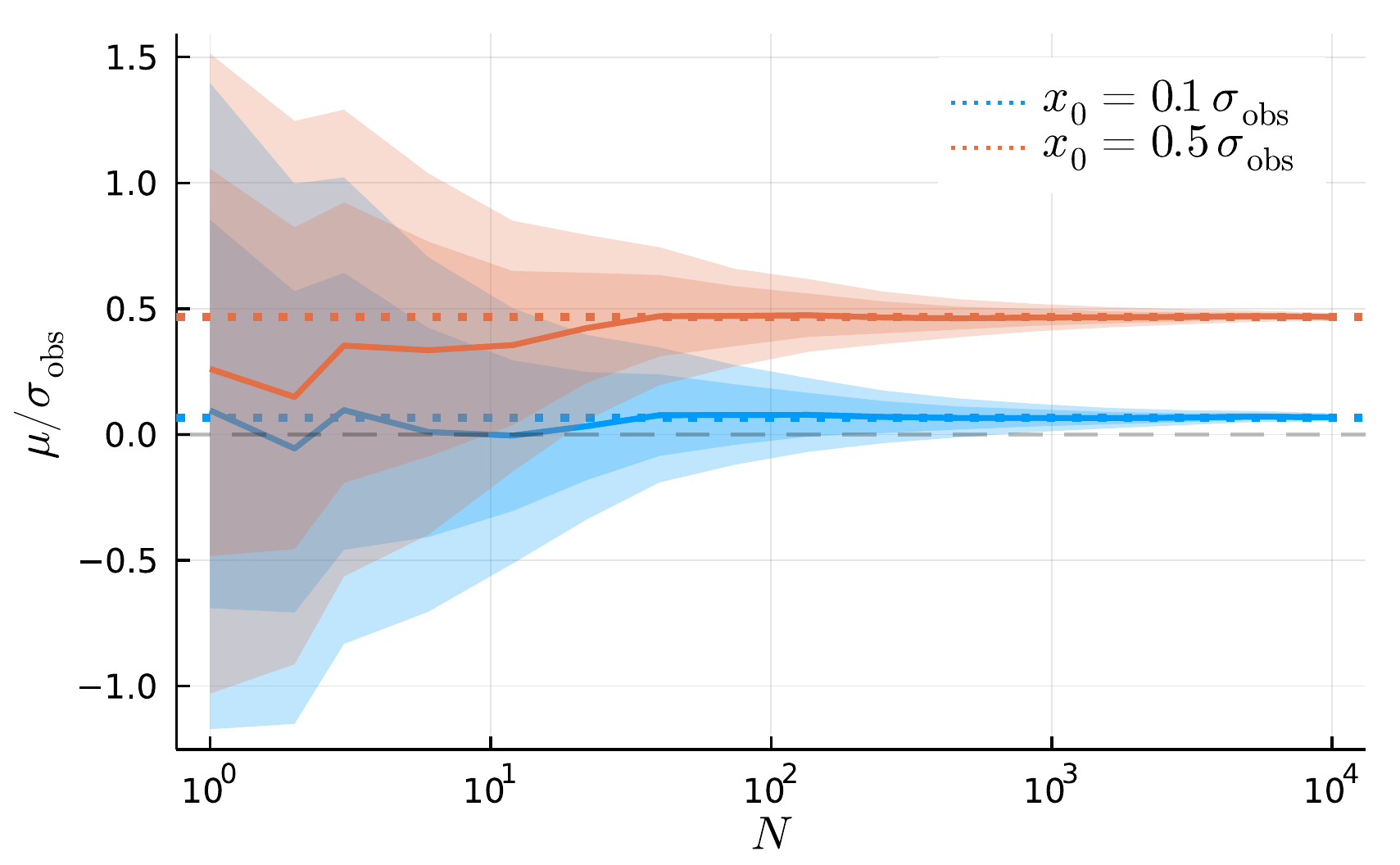}
  \includegraphics[width=\columnwidth]{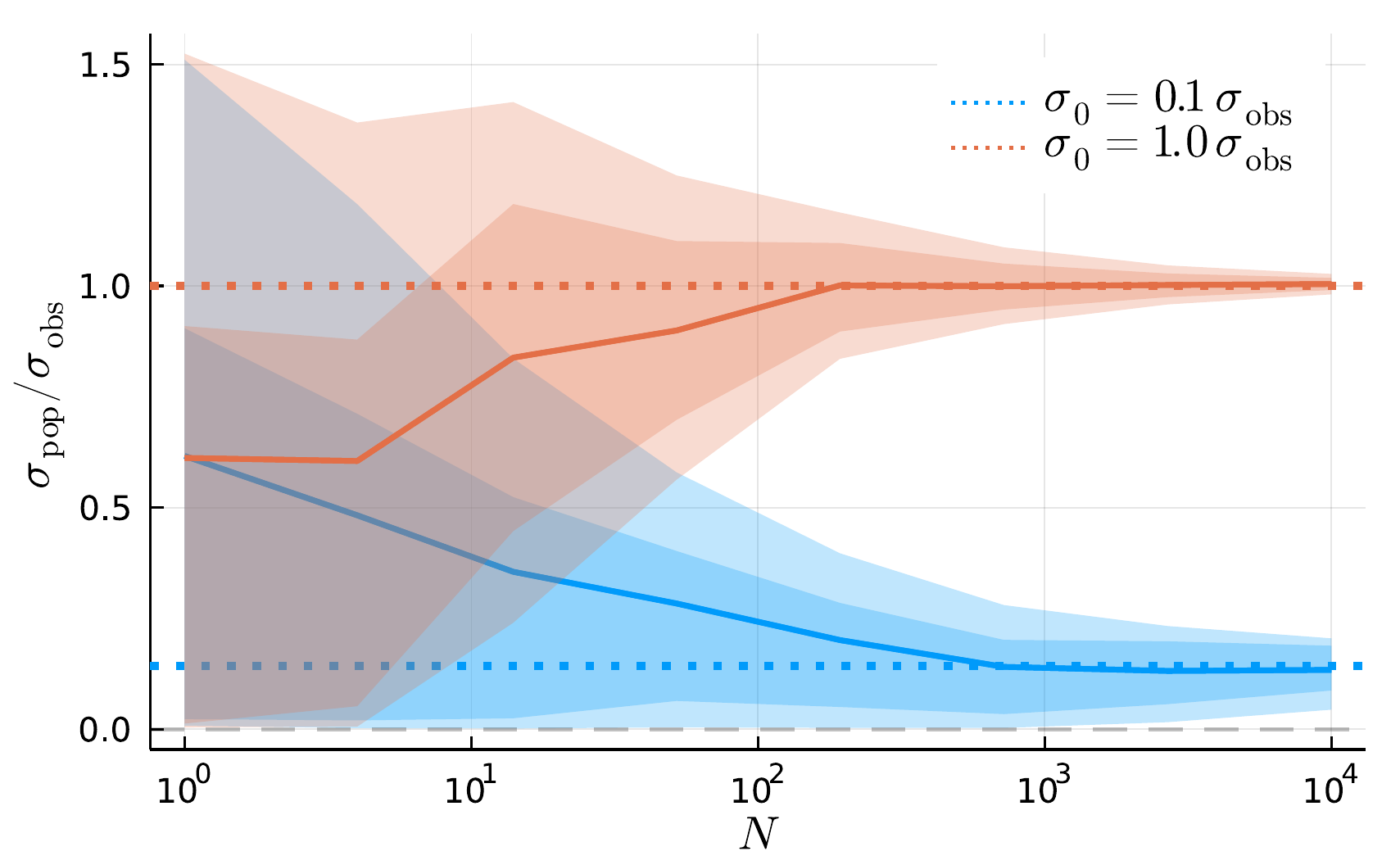}
  \caption{\label{fig:toy-musigma-1d} Recovered population mean ($\mupop$, left) and standard deviation ($\sigmapop$, right) as a function of catalog size ($N$, abscissa), for our two toy models: a fixed deviation $\xtrue_0$ for all events (left) and a deviation scatter $\sigma_0$ across events (right).
  In each case, we show two values for the true deviation: $\xtrue_0, \sigma_0 = 0.1\, \sigmaobs$ (blue) and $\xtrue_0, \sigma_0 = 0.8\, \sigmaobs$ (red).
  The measurement is represented by posterior median (solid lines) surrounded by 68\% and 90\% highest-density credible bands (shading); we also show the true value (dotted lines) and the null expectation (dashed gray line).
  Smaller deviations require more observations to be detected---for example, we only need $N \gtrsim 3$ events to notice $\mupop > 0$ at $1\sigma$ (68\% credibility) if $\xtrue_0=0.8\,\sigmaobs$, but $N\gtrsim 100$ if $\xtrue_0 = 0.1\, \sigmaobs$.
  }
\end{figure*}

Besides indicating that the data are inconsistent with the null hypothesis, the hierarchical analysis provides descriptive information about the nature of the deviation.
With enough observations, the measurements of $\mupop$ and $\sigmapop$ converge to the $\xtrue_0$ and $\sigma_0$ values respectively for the two models with increasing precision for larger $N$.
In Fig.~\ref{fig:toy-musigma-1d}, we show this behavior explicitly for two example magnitudes of the GR deviation.
As expected, we can detect larger deviations with fewer detections, and need more observations to notice a nonzero scatter than a nonzero mean.

Prior choices play a lesser role in the hierarchical approach than in BF computations.
The hierarchical analysis takes as input the likelihoods for individual events, so the prior used to initially analyze the data is largely irrelevant as long as it offers the likelihood ample support (e.g., $\priorw \gg \sigmaobs$ in the notation of the previous section).
The choice of prior is, instead, transferred to the $\mupop$ and $\sigmapop$ hyperparameters; however, any reasonably smooth choice will work assuming we have a enough events for the hierarchical measurement to be informative.
If observations are not sufficiently numerous, the result will be influenced by the $\mupop$ and $\sigmapop$ prior.

\begin{figure*}
  \includegraphics[width=\columnwidth]{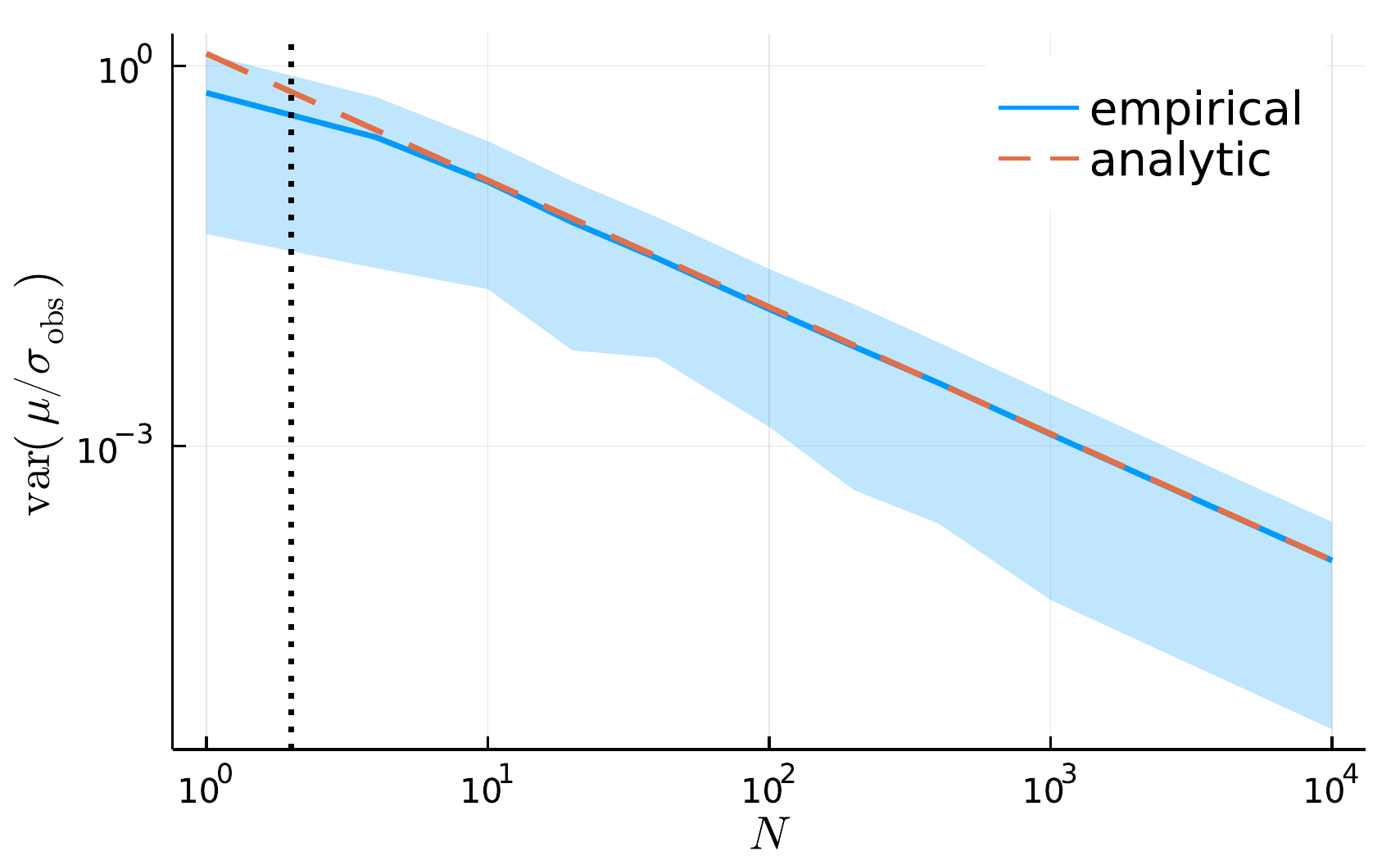}
  \includegraphics[width=\columnwidth]{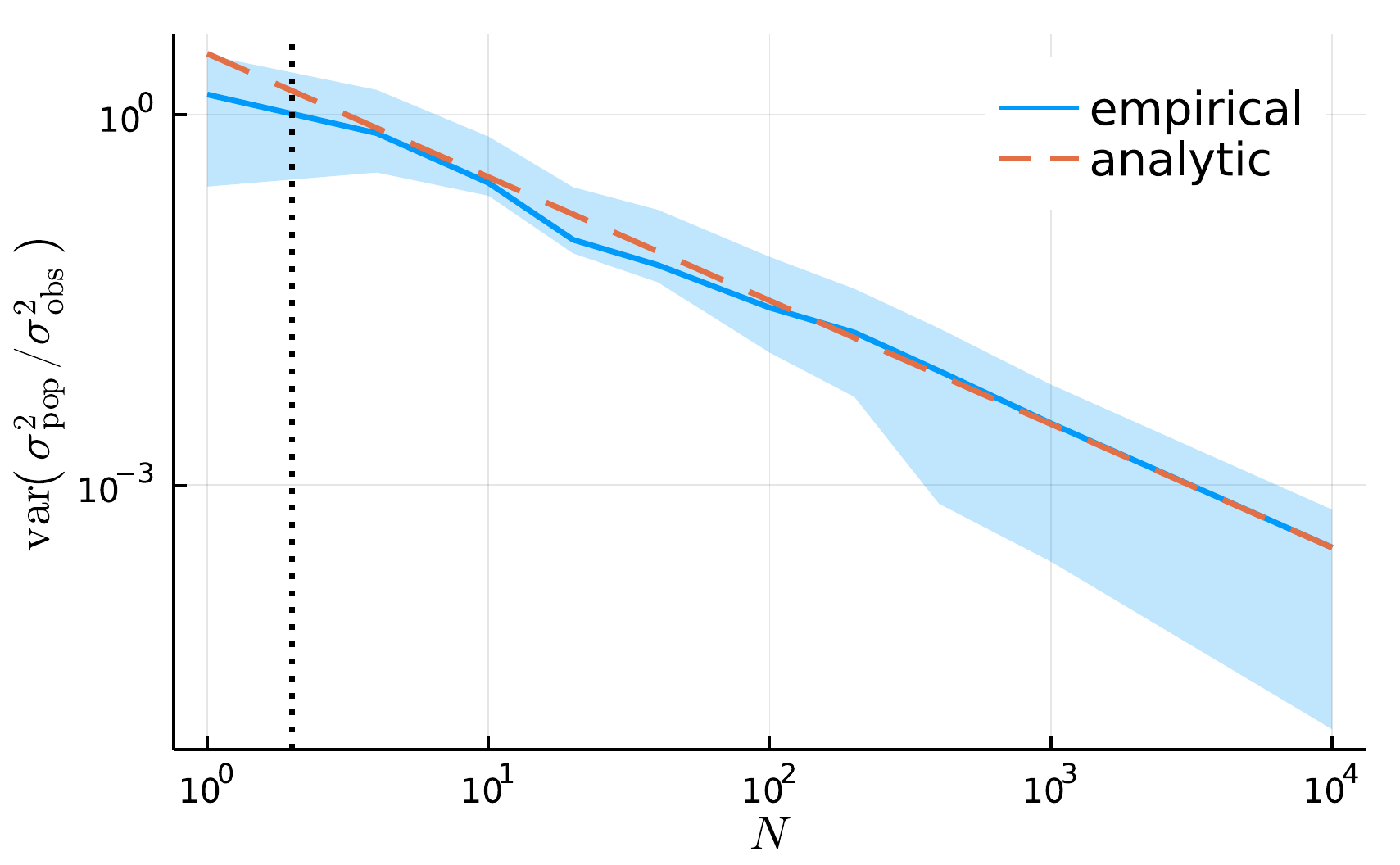}
  \caption{\label{fig:hier-scaling} Scaling of the uncertainty in the recovered hyperparameters $\mupop$ (left) and $\sigmapop^2$ (right) as a function of catalog size $N$, computed for a case in which the true values are $\mupop=0$ and $\sigmapop=\sigmaobs/2$.
  A solid line marks the median of the posterior variance computed over 50 simulated catalogs for any given $N$, while surrounding bands enclose the corresponding 68\% highest-density interval.
  For low $N$, the variances $\var\left(\mu/\sigmaobs\right)$ and $\var\left(\sigmapop^2/\sigmaobs^2\right)$ are dominated by the hyperprior; as $N$ increases, the data become more informative and the variances approach the analytic prediction of Eqs.~\eqref{eq:hier-varmu} and \eqref{eq:hier-varsigma} (red dashed lines).
  Dotted vertical lines mark the expected number of observations needed for the likelihood of Eq.~\eqref{eq:hier-like} to become narrower than the prior, as dictated by Eqs.~\eqref{eq:hier-thresh-mu} and \eqref{eq:hier-thresh-sigma} with $\sigmaprior = \sigmaobs$, but rounded up to the closest integer.
  }
\end{figure*}

In the case of our idealized examples, we can analytically predict the number of detections needed for the hierarchical measurement to be informative.
As we show in Appendix \ref{app:hier}, the variance of the marginalized hierarchical likelihood is expected to scale as
\begin{equation} \label{eq:hier-varmu}
\var\left({\mu}\right) = \frac{\sigmaobs^2 + \sigma_0^2}{N}
\end{equation}
for the inferred population mean and
\begin{equation} \label{eq:hier-varsigma}
\var\left({\sigmapop^2}\right) = \frac{2}{N} \left(\sigmaobs^2 + \sigma_0^2\right)^2
\end{equation}
for the inferred population variance, assuming the true population variance is $\sigma_0^2$ and irrespective of the true mean.
As a rule of thumb, the hierarchical measurement will become informative once the characteristic width of the likelihood of Eq.~\eqref{eq:hier-like} becomes smaller than the scale imposed by the prior.
With hyperpriors of scale $\sigmaprior$, this implies that the $\mupop$ measurement should start becoming informative (in the sense that we obtain a hyperposterior narrower than the prior) once we accumulate
\begin{equation} \label{eq:hier-thresh-mu}
N \gtrsim \frac{\sigmaobs^2 + \sigmapop^2}{\sigmaprior^2} = \left(\frac{\sigmatot}{\sigmaprior}\right)^2
\end{equation}
measurements; meanwhile, for $\sigmapop$, the equivalent requirement is
\begin{equation} \label{eq:hier-thresh-sigma}
N \gtrsim \frac{\left(\sigmaobs^2 + \sigmapop^2\right)^2}{\sigmaprior^4}
= \left(\frac{\sigmatot}{\sigmaprior}\right)^4,
\end{equation}
as we show in Appendix \ref{app:hier}.  

\begin{figure*}
  \includegraphics[width=\columnwidth]{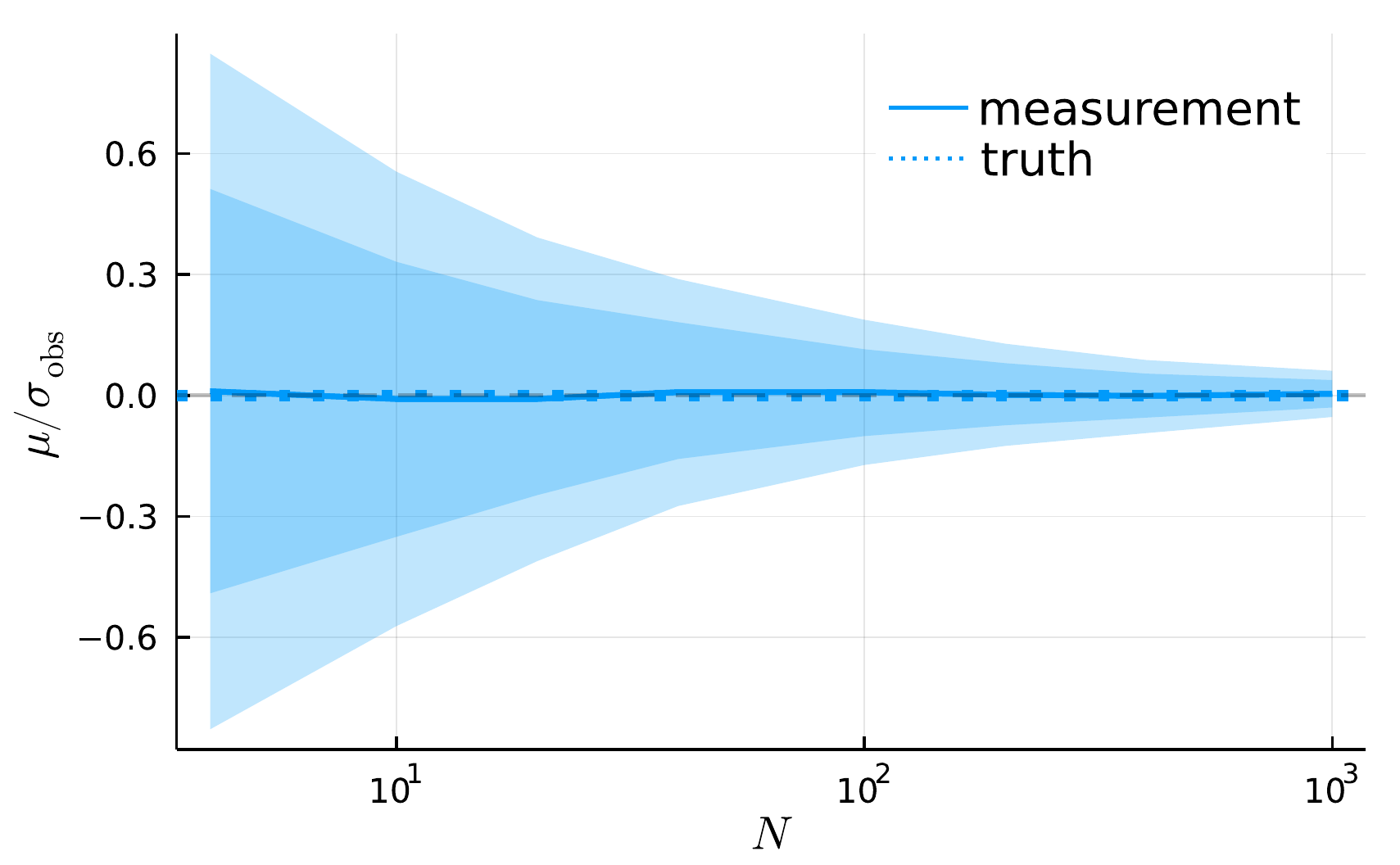}
  \includegraphics[width=\columnwidth]{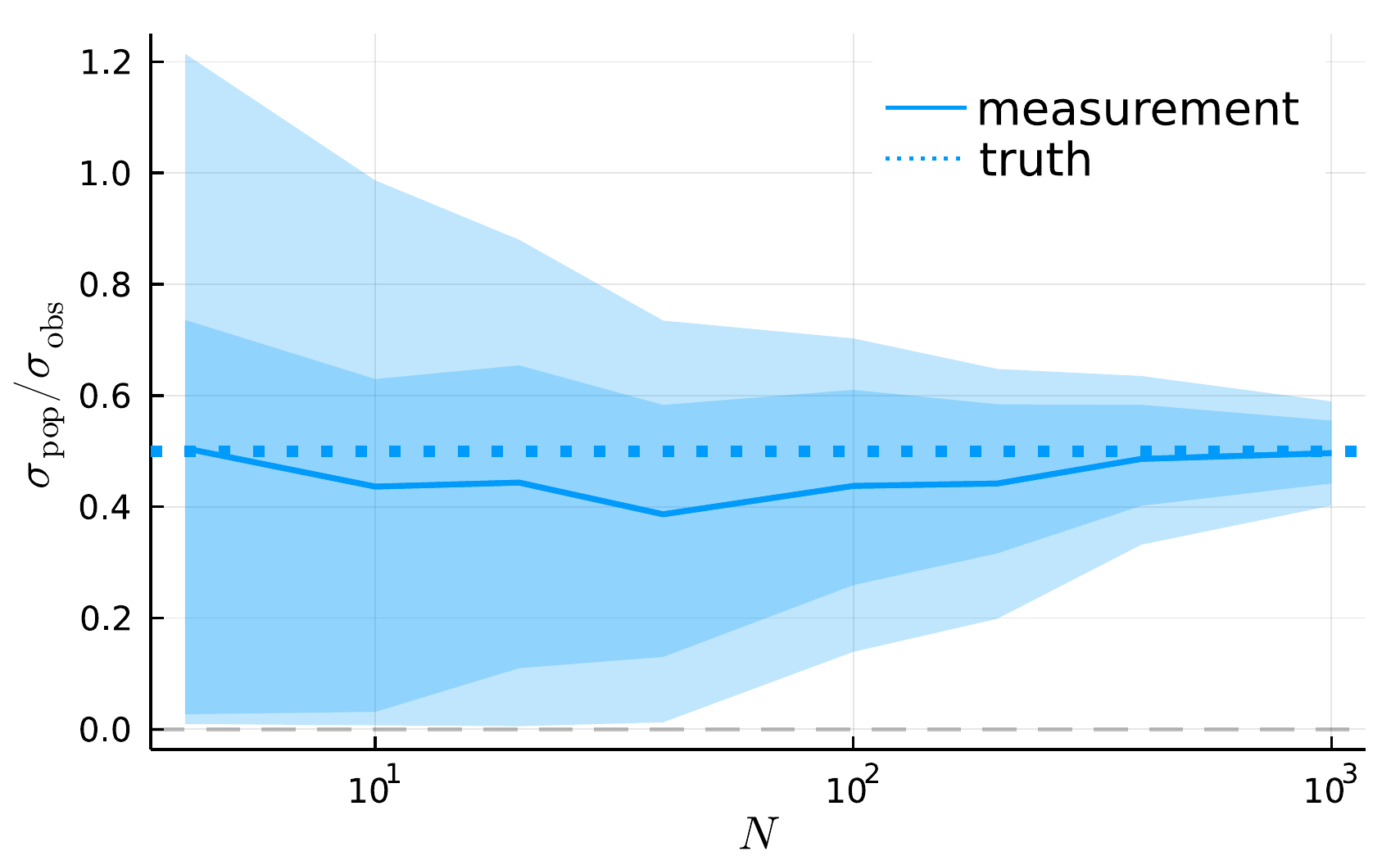}
  \caption{\label{fig:hier-fencepost-musigma} As in Fig.~\ref{fig:toy-musigma-1d}, except the true population follows the fencepost model of Sec.~\ref{sec:fencepost}, by which the true deviation is $\xtrue = \pm \sigmaobs / 2$ with equal probability for either sign. Even though this distribution cannot be expressed as the limit of a Gaussian, the hierarchical analysis infers the correct values of $\mupop = 0$ and $\sigmapop = \sigmaobs/2$.
  }
\end{figure*}

We demonstrate these scalings in Fig.~\ref{fig:hier-scaling}, where we show the variance in the inferred population mean and variance from simulated populations of measurements with $\sigma_0 = \sigmaobs/2$ and no mean, and setting $\sigmaprior = \sigmaobs$ for concreteness.
For increasing number of measurements $N$, the posterior converges to the true values ($\mupop=0, \sigmapop = \sigmaobs/2$).
For small catalogs, i.e., for $N$ comparable or smaller than the thresholds above, the average uncertainty in these measurements is broad but smaller than expected simply from Eqs.~\eqref{eq:hier-varmu} and \eqref{eq:hier-varsigma} because it is dominated by the prior.
As the number of detections increases, the posterior variance becomes well described by Eqs.~\eqref{eq:hier-varmu} and \eqref{eq:hier-varsigma}, meaning that the data become informative and the likelihood dominates over the prior.

Alternately, we may ask how many measurements would be required to establish a
non-vanishing population mean or variance.  This requires $\var\left( \mu
\right) \lesssim x_0^2$ or $\var \left( \sigmapop^2 \right) \lesssim
\sigma_0^4$.  The former would imply 
\begin{equation}
  N \gtrsim \left( \frac{\sigmatot}{x_0} \right)^2
\end{equation}
and the latter 
\begin{equation}
  N \gtrsim 2 \left( \frac{\sigmatot}{\sigma_0} \right)^4.
\end{equation}

\section{Fencepost model}
\label{sec:fencepost}

In the examples above, the simulated populations could be perfectly reproduced as special cases of the hierarchical population model---that is, there existed a choice of $\mupop$ and $\sigmapop$ for which the hierarchical model reduced exactly to the true distribution we simulated.
Of course, we do not necessarily expect this to be the case in reality: a deviation from GR could manifest as a nontrivial function of the source parameters and the coupling constants in the theory; similarly, for other effects such as memory, it is not realistic to expect the true population of parameters to be fully described by a simple Gaussian if the null hypothesis is incorrect.
In light of that, one might worry that the hierarchical method only outperformed BFs in the above examples because the hierarchical model was able to match the true population exactly, and that this gain would fail to materialize in realistic situations.
Yet, as we show in this section, this is not the case: the hierarchical method is more robust than products of BFs even when the underlying population cannot be fit exactly by a Gaussian.

Consider a situation in which the true deviation parameter is either $\xtrue=\pm \xtrue_0$ for some $\xtrue_0$ and with equal probability for both signs.
The true distribution in this ``fencepost'' model is simply the sum of two delta functions,
\begin{equation}
p_3\left(\xtrue\right) = \frac{1}{2}\left[ \delta\left(\xtrue - \xtrue_0\right) + \delta\left(\xtrue + \xtrue_0\right) \right] ,
\end{equation}
and, therefore, has zero mean and standard deviation $\sigma_0 = |\xtrue_0|$.
This population cannot be described as a limiting case of a Gaussian distribution; nevertheless, we can always analyze it hierarchically with the Gaussian likelihood of Eq.~\eqref{eq:hier-like}, and expect to recover the correct values for the population mean and spread (namely, $\mupop = 0$ and $\sigmapop = |\xtrue_0|$) given enough detections~\cite{Isi:2019asy}.
In Fig.~\ref{fig:hier-fencepost-musigma}, we show this explicitly for simulated measurements in which $\xtrue_0 = \sigmaobs/2$.

\begin{figure}
  \includegraphics[width=\columnwidth]{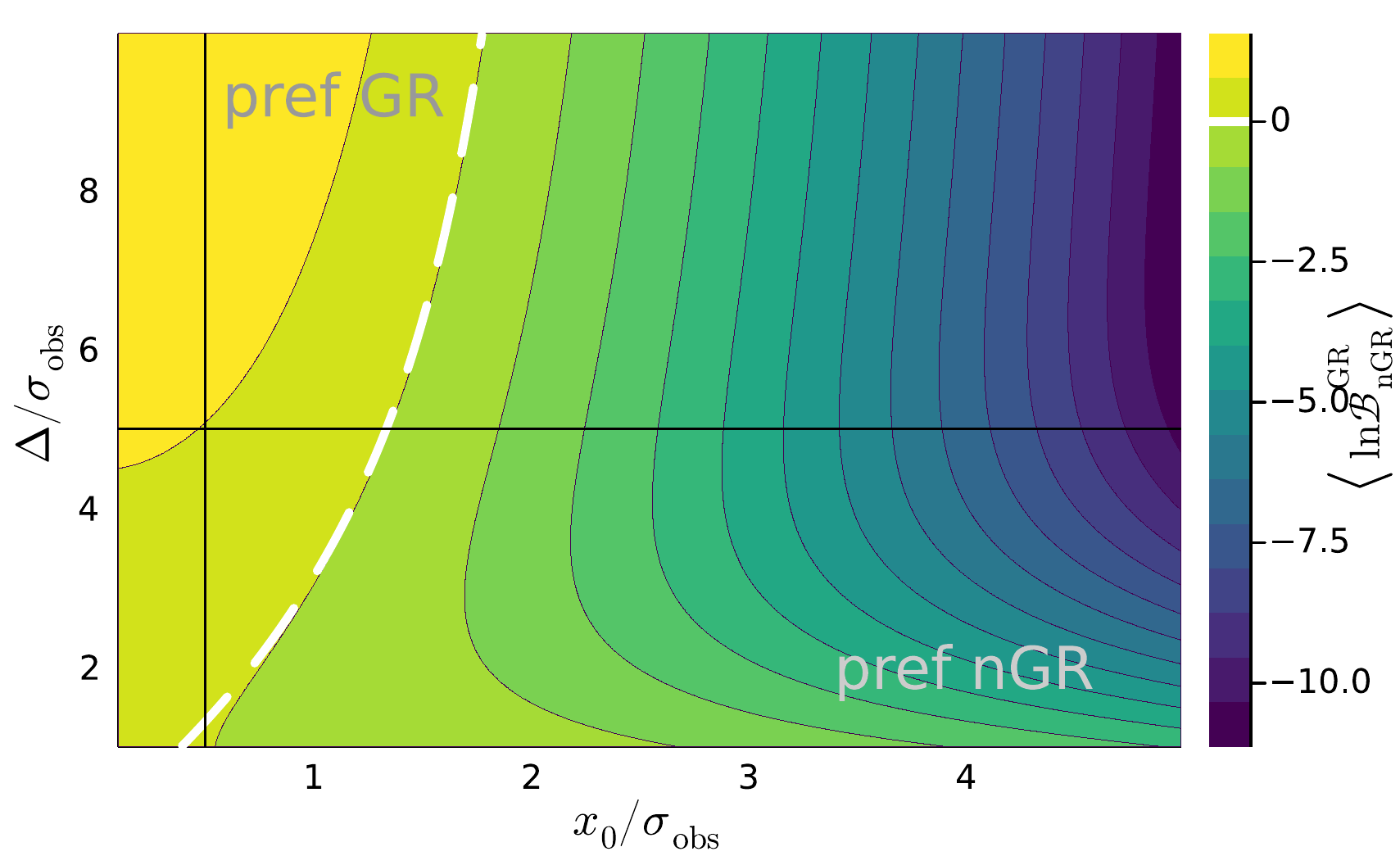}
  \caption{\label{fig:fencepost-lnZ} Average log BF in favor of GR accumulated for each observation in the ``fencepost'' model described in Sec.~\ref{sec:fencepost} as a function of the prior width $\priorw$ and the location of the true deviations at $\pm \xtrue_0$.  The black lines indicate the parameter values chosen for the numerical experiment illustrated in Fig.~\ref{fig:fencepost-expt}.}
\end{figure}

On the other hand, when presented with the fencepost model, BF computations suffer from the same problems already identified above: with a broad prior relative to the measurement uncertainty ($\priorw > \sigmaobs$), the combined BF for multiple observations will necessarily converge to the wrong answer (i.e., favor of the null hypothesis) unless $\xtrue_0 \gtrsim \sigmaobs$, in which case the deviation is detectable in individual observations.
Similarly to Fig.~\ref{fig:toy-BF-parameter-scan}, in Fig.~\ref{fig:fencepost-lnZ} we show the scaling of the expected BF accumulated from fencepost-model observations, as a function of $x_0$ and $\priorw$.

\begin{figure*}
  \includegraphics[width=\columnwidth]{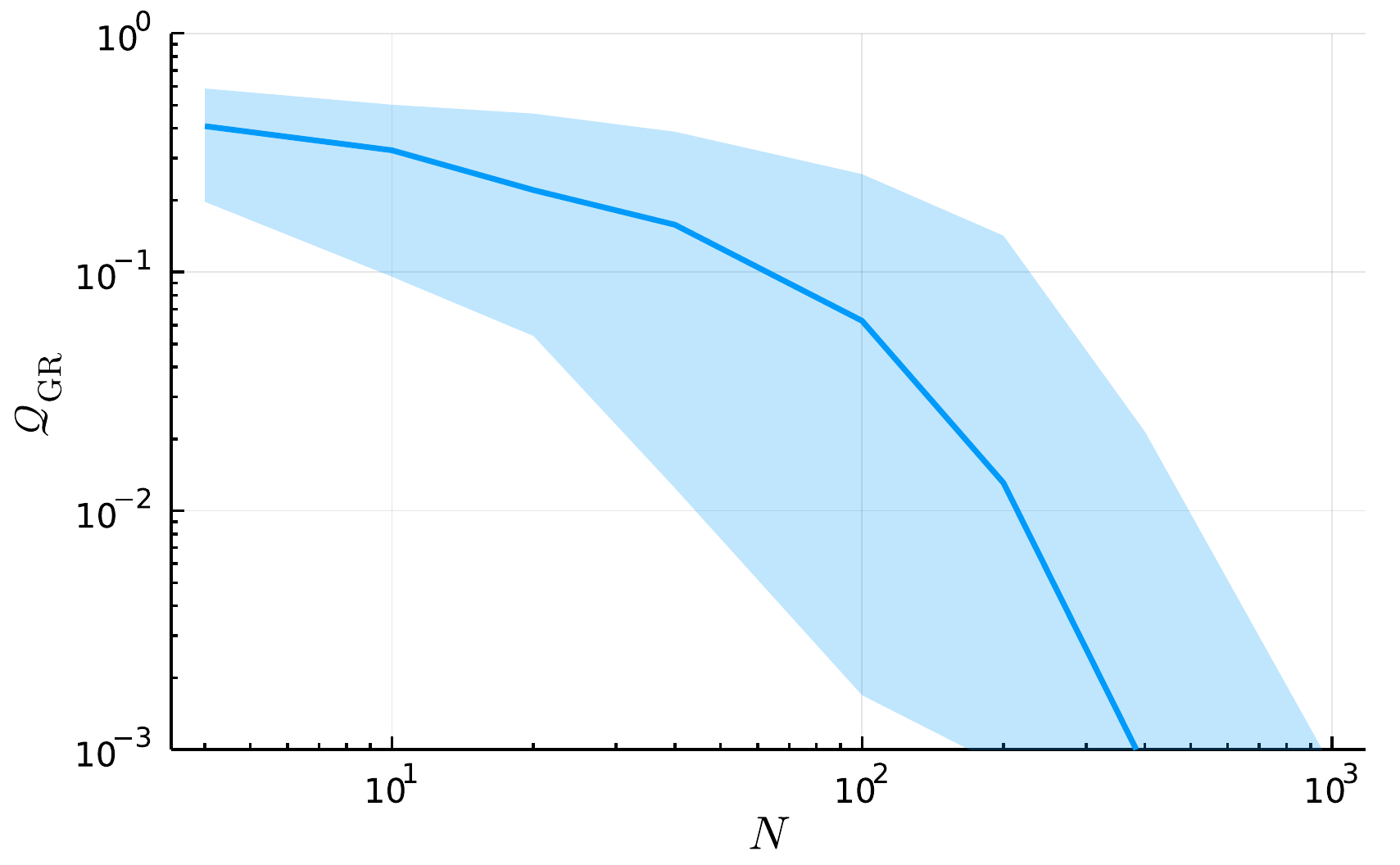}
  \includegraphics[width=\columnwidth]{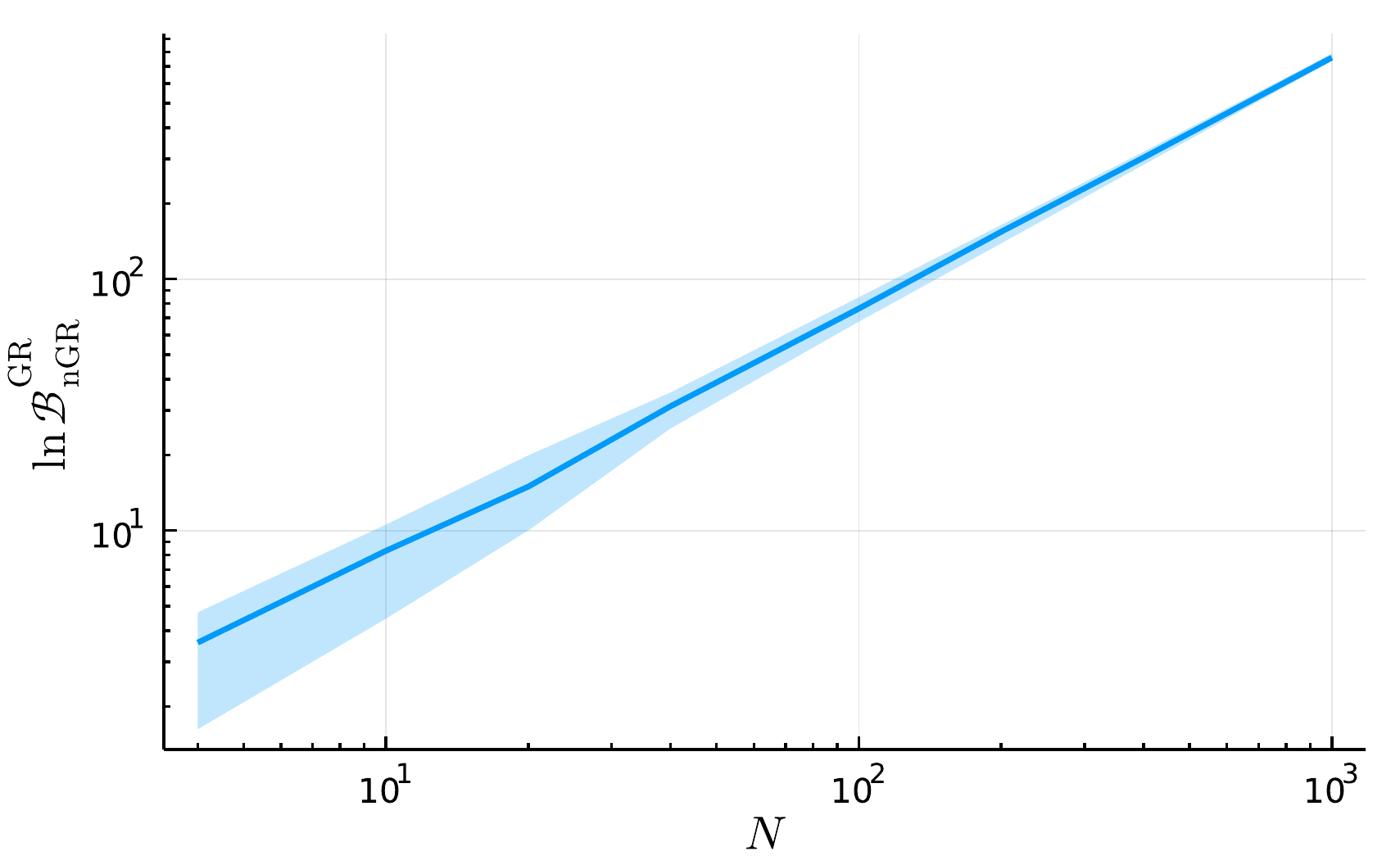}
  \caption{\label{fig:fencepost-expt} Accumulation of credibility (left) or log
  BF (right) for GR versus catalog size $N$ in numerical experiments
  corresponding to the parameter choices indicated by the black lines in Fig.~\ref{fig:fencepost-lnZ}.  The solid line gives the median over 100 catalog
  realizations at each catalog size $N$ while the band shows the range of the
  16th to 84th percentile values.  The credibility of GR (left) is the fraction
  of posterior mass for $\mu$ and $\sigma$ in our hierarchical model that lies
  at a lower posterior density than the GR values $\mu = \sigma = 0$ when the
  model is fit to a catalog of observations whose true and observed deviations
  are drawn from the ``fencepost'' model described in Sec.~\ref{sec:fencepost} with $\xtrue = \pm \sigma_\mathrm{obs}/2$.  The log Bayes
  factor (right) is the sum of the log BFs for each observation in the
  catalog using a flat prior on the true deviation $-\priorw < \xtrue < \priorw$
  with $\priorw = 5 \sigma_\mathrm{obs}$.  The hierarchical model correctly finds
  that there is little credibility for GR once the catalog size is a few
  hundred; the accumulated BF, on the other hand, becomes very
  confident in the incorrect GR model even at small catalog sizes.}
\end{figure*}

In Fig.~\ref{fig:fencepost-expt}, we compare the hierarchical and Bayes-factor results for a progressively-higher number of simulated measurements from a fencepost population with $\xtrue_0 = \sigmaobs/2$.
As the number of measurements increases, the hierarchical model disfavors the null hypothesis more strongly, with $\qgr$ vanishing rapidly for increasing $N$.
On the other hand, a BF computed with $\priorw = 5\sigmaobs$ grows exponentially in favor of the null hypothesis, yielding a spectacularly incorrect result.

\section{Conclusions}
\label{sec:conclusion}

Although conceptually appealing in idealized situations, the use of BFs to
aggregate information from multiple observations presents difficulties in
practice. Their apparent simplicity in reducing a complex model selection
problem to a single number hides an opaque dependence strict and unrealistic
population assumptions. Unless priors (aka, the ``population'') adapt to the
observations at hand, BFs are difficult to interpret---a problem that is
compounded when multiplying such BFs from a catalog of observations. Even when
priors are adequate, the result on its own provides no insight as to \emph{why}
a model is to be preferred over another. This and related problems have been
widely discussed in the statistics literature \cite[e.g.,][and references
therein]{Lotfi2022}, but not extensively in the context of GWs and testing GR.

In this paper, we have examined BFs and hierarchical posteriors as two commonly-used alternatives to derive information from collections of GW detections in order to decide between two models, e.g., the presence or absence of beyond-GR effects in the detected waveforms.
We furthered arguments in~\cite{Zimmerman:2019wzo,Isi:2019asy} to show that, without a principled way to set priors, BFs are an unreliable tool for this task.
We demonstrated this with three examples in which the value of some parameter $x$ encoding the effect in question (e.g., a deviation from GR) follows different distributions deviating from the null hypothesis.

We have found that, when the truth does not conform to the null model, the usual approach of multiplying single-event BF converges to the incorrect answer for an increasing number of observations, except in a regime where the targeted effect is discernible in individual observations (thus negating the need for combining events in the first place).  
On the other hand, hierarchical modeling of the underlying population leads to identification of the appropriate priors (aka, the ``population model") and converges to the correct answer.  We established this in the context of nested models for which GR can be recovered as a special case of the beyond-GR model (i.e., $x_0 = 0$); however, the issue of sensitivity to the prior width will still be present in non-nested models \cite[e.g.,][]{Laghi2021} where it will require a different solution.

In principle, BFs could be computed after the hierarchical population
inference~\cite{Lotfi2022} or between different population
models~\cite{LIGOScientific:2020kqk}, but we here show that they are unreliable
without this step. Even then, it is not possible to evade the core problem of
prior dependence when computing BFs, no matter how many levels of inference are
applied: the BF computation based on the highest level of inference in a
hierarchical model will still depend on the choice of priors on that level,
reducing the problem once again to the choice of a prior distribution that is
difficult to establish in a principled way.  This issue is devistatingly acute
for the approach that multiplies Bayes factors with a simple, fixed prior
because each observation contributes an additional prior factor.

\begin{acknowledgments}
We are grateful to Tyson Littenberg for insightful feedback on this manuscript.
During part of this work, M.I.\ was supported by NASA through the NASA Hubble Fellowship
grant No.\ HST-HF2-51410.001-A awarded by the Space Telescope
Science Institute, which is operated by the Association of Universities
for Research in Astronomy, Inc., for NASA, under contract NAS5-26555.
The Flatiron Institute is a division of the Simons Foundation, supported through the generosity of Marilyn and Jim Simons.  We thank Gregorio Carullo for comments on non-nested models.
This paper carries LIGO document number \dcc{}.
Software: {\tt matplotlib}~\cite{Hunter:2007}, {\tt julia}~\cite{bezanson2012julia}, {\tt Turing.jl} \cite{ge2018t}, {\tt Plots.jl} \cite{tom_breloff_2022_6365416}.
\end{acknowledgments}

\appendix
\section{Expectation value and variance of hierarchical parameters}
\label{app:hier}

Consider $i = 0\dots N-1$ measurements of parameters $x_i$ whose true values, $\mu_i$, are drawn from a normal distribution $\mu_i \sim \mathcal{N}(\mupop,\,\sigmapop)$; further assume each measurement is unbiased, i.e., $\langle x_i \rangle = \mu_i$, where the angle brackets denote a noise average, and that it is well represented by a Gaussian distribution such that $x_i \sim \mathcal{N}(\mu_i,\, \sigma_i)$, where $\sigma_i$ is the measurement uncertainty.%
\footnote{In the main text, we used slightly different notation: instead of $\left(x_i, \mu_i, \sigma_i\right)$ we had $\left(\xobs, \xtrue, \sigmaobs\right)$; the former is slightly more succinct, which will be helpful here given the increased number of mathematical expressions.}

The joint likelihood for $\mupop$ and $\sigmapop$ conditional on the $N$ uncertainties $\sigma_i$ can be
obtained by marginalizing over the true values $\mu_i$:
\begin{align}
p(x_i \mid \mupop, \sigmapop, \sigma_i) &= \int p(x_i \mid \mu_i,\sigma_i)\, p(\mu_i \mid \mupop, \sigmapop) \infd \mu_i \nonumber\\
&= \frac{1}{2\pi\sigma_i\sigmapop} \int e^{-\frac{\left(x_i -\mu_i\right)^2}{2\sigma_i^2}} e^{-\frac{\left(\mu_i - \mupop\right)^2}{2\sigmapop^2}} \infd \mu_i \nonumber\\
&= \frac{1}{\sqrt{2\pi \sigma^2_{\mathrm{tot},i}}} e^{-\frac{\left(x_i - \mupop\right)^2}{2\sigma_{\mathrm{tot},i}^2}} \, ,
\end{align}
where $\sigma_{\mathrm{tot},i}^2 = \sigma_i^2 + \sigmapop^2$ is the total variance for the $i$th measurement.
For the full set of $N$ measurements $\{x_i\}$, then, the hierarchical likelihood is
\begin{equation}
p( \{x_i\} \mid \mupop, \sigmapop, \{\sigma_i\}) = \prod_{i=0}^{N-1} \frac{1}{\sqrt{2\pi \sigma_{\mathrm{tot},i}^2}} e^{-\frac{\left(x_i - \mupop\right)^2}{2\sigma_{\mathrm{tot},i}^2}}.
\end{equation}
The maximum-likelihood estimators for $\mupop$ and $\sigmapop$, denoted $\mupopest$ and $\sigmapopest$ respectively, can be found by
enforcing
\begin{equation}
\left. \partial_{\mupop} \ln \left[ p( \{x_i\} \mid \mupop, \sigmapop, \{\sigma_i\}) \right] \right|_{\mupop=\mupopest} = 0 \,,
\end{equation}
where $\partial_{\mupop}$ denotes the partial derivative $\frac{\partial}{\partial \mupop}$, and similar for $\sigmapop$. We find that $\mupopest$ and $\sigmapopest$ satisfy
\begin{equation}
\mupopest = \frac{\sum x_i w_i}{\sum w_i}\, ,
\end{equation}
and
\begin{equation}
\sum w_i - \sum w_i^2 \left(x_i - \mupopest\right)^2 = 0
\end{equation}
for $w_i \equiv \sigma_{\mathrm{tot},i}^{-2} = \left(\sigma_i^2 + \sigmapopest^2\right)^{-1}$.

We cannot solve this most general (heteroskedastic) case for $\mupopest$ and $\sigmapopest$ in closed form, so we specialize to the (homoskedastic) case where all measurements have similar error, $\sigma_i = \sigmaobs$ for all $i$.
With this simplification, the above relations reduce to
\begin{equation}
\mupopest = \frac{1}{N} \sum x_i\, ,
\end{equation}
and
\begin{equation}
\sigmapopest^2 = \frac{1}{N} \sum \left(x_i - \mupopest\right)^2 - \sigmaobs^2 \, .
\end{equation}
As might be expected, the maximum-likelihood estimate of the population mean is simply the sample mean, and the inferred population variance corresponds to the variance in the data that cannot be accounted for by the statistical uncertainty in each individual measurement.

We can go one step further and compute the uncertainty in these estimators, which we can take as a proxy for the width of the marginal likelihoods.
The uncertainty in $\mupopest$ is straightforward to compute, since this is just a linear combination of independent random variables $x_i$ with known variance $\sigmatot^2 \equiv \sigmaobs^2 + \sigmapop^2$, hence
\begin{equation} \label{eq:hier-varmu-app}
\var \left(\mupopest\right) = \frac{\sigmaobs^2 + \sigmapop^2}{N}\, .
\end{equation}
Obtaining $\var \left(\sigmapopest^2\right)$ is less straightforward, but we can do so by writing $\var \left(\sigmapopest^2\right) = F^{-1}$, in terms of the corresponding Fisher element%
\footnote{We carry out this calculation for $\sigmapop^2$ instead of $\sigmapop$ because the latter is irregular at the origin. }
\begin{equation}
F \equiv - \left\langle \left.\partial^2_{\sigmapop^2} \ln \left[ p( \{x_i\} \mid \mupop, \sigmapop, \sigmaobs) \right] \right|_{\sigmapop^2=\sigmapopest^2} \right\rangle .
\end{equation}
Doing the math, we find
\begin{equation} \label{eq:hier-varsigma-app}
\var \left(\sigmapopest^2\right) = 2 \frac{\left(\sigmaobs^2 + \sigmapopest^2\right)^2}{N}\ .
\end{equation}
In the main text, we quoted these results for $\var \left(\mupopest\right)$ and $\var \left(\sigmapopest^2\right)$ in Eqs.~\eqref{eq:hier-varmu} and \eqref{eq:hier-varsigma} respectively.

We can compare the width of the hierarchical likelihood, as proxied by the estimator variances above, to some typical scale of interest in the problem. 
Below we consider the scale imposed by the $\mupop$ and $\sigmapop$ hyperpriors to estimate the number of events before the likelihood become informative 
with respect to the prior.
We chose the hyperpriors to be Gaussians with scale $\sigmaprior$, restricting to positive $\sigmapop$ values, i.e.,
\begin{equation}
p(\mupop\mid \sigmaprior) = \frac{1}{\sqrt{2\pi\sigmaprior^2}} \exp\left(-\frac{\mu^2 }{2\sigmaprior^2}\right)
\end{equation}
for the mean, and 
\begin{equation}
p(\sigmapop\mid \sigmaprior) =
\begin{cases}
\sqrt{\frac{2}{\pi\sigmaprior^2}}\exp\left(-\frac{\sigmapop^2 }{2\sigmaprior^2}\right) \hfill (\sigmapop \geq 0) \\%
0 \hfill (\sigmapop < 0)
\end{cases}
\end{equation}
for the standard deviation, where the difference in normalization arises from the $\sigmapop \geq 0$ truncation.
The prior variances in our example are, thus, $\var\left(\mupop\right) = \sigmaprior^2$ for the mean, and $\var\left(\sigmapop^2\right) = 2\sigmaprior^4$ for the variance (obtained through direct computation).
For concreteness, in the main text we set $\sigmaprior = \sigmaobs$, since that is the only scale intrinsic to the measurement.

We can now directly compute the number of observations required for the likelihood to achieve comparable widths by equating these variances to the likelihood variances from Eqs.~\eqref{eq:hier-varmu-app} and \eqref{eq:hier-varsigma-app} above.
The result is
\begin{equation}
N \gtrsim \frac{\sigmaobs^2 + \sigmapop^2}{\sigmaprior^2}
\end{equation}
for $\mupop$, and
\begin{equation} \label{eq:hier-thresh-sigma-app}
N \gtrsim \left(\frac{\sigmaobs^2 + \sigmapop^2}{\sigmaprior^2}\right)^2
= \left(\frac{\sigmatot}{\sigmaprior}\right)^4
\end{equation}
for $\sigmapop$.
We quoted these results in Eqs.~\eqref{eq:hier-thresh-mu} and \eqref{eq:hier-thresh-sigma} in the main text.
We require at least this many measurements before the uncertainty in the population variance can be smaller than the measurement uncertainty.
Until that point, the $\sigmapop$ posterior will be dominated by the prior.

The results for the $N$ thresholds quoted above hinge on the specific choice of prior for $\mupop$ and $\sigmapop$.
In the main text, we justified our decision to set the scale of those priors based on $\sigmaobs$ by noting that this is the only intrinsic scale to the problem, and should always be sufficiently broad as long as the deviation from GR is not visible in a single detection---the regime in which we are interested in the first place.
Had we chosen to increase the prior variance by some factor, then the $N$ thresholds would decrease by the same factor, i.e., we need fewer detections to gain information relative to a broad (less informative) prior than a narrower prior.
Either way, the result converges to the right answer as we accumulate more observations.

\bibliography{cbc-group}

\end{document}